\documentclass{aa}
\usepackage{natbib,graphics,txfonts}

\begin{document}

\title{Investigating early-type galaxy evolution with a multiwavelength approach
\footnote{Tables 4 are available in electronic form
at the CDS via anonymous ftp to cdsarc.u-strasbg.fr (130.79.128.5)
or via http://cdsweb.u-strasbg.fr/cgi-bin/qcat?J/A+A/}}
\subtitle{II.  The UV structure of 11 galaxies with {\it Swift}-{\tt UVOT}} 

\author{
R. Rampazzo\inst{1}, 
P. Mazzei\inst{1},
A. Marino\inst{1},
M. Uslenghi\inst{2},
G. Trinchieri\inst{3},
A. Wolter\inst{3}
} 

\institute{INAF Osservatorio Astronomico di Padova, Vicolo
dell'Osservatorio 5, 35122 Padova, Italy\\
\email{roberto.rampazzo@oapd.inaf.it}
\and
INAF-IASF, via E. Bassini 15, 20133 Milano, Italy
\and
INAF Osservatorio Astronomico di Brera, Via Brera 28, 20121 Milano, Italy
}
\titlerunning{Investigating ETGs evolution with a multi-wavelength approach}
  \authorrunning{Rampazzo et al.}
  \date{Received; accepted}

\abstract
{{\it GALEX} detected  a  significant fraction of early-type galaxies, 
in particular S0s, showing Far-UV bright structures, sometimes involving an
entire galaxy out to its outskirts. These features suggest the presence of either 
recent, ongoing and/or prolonged star formation episodes, shedding new light on the 
evolution of these systems.}
{We aim at understanding the evolutionary path[s] of these early-type galaxies and 
the mechanisms at the origin of their UV-bright structures. We investigate with a
multi$\lambda$ approach the link between the inner and the 
outer galaxy regions of a set of eleven  early-type galaxies
selected because of their nearly passive stage of evolution in the nuclear region.}
{This paper, second of a series, focuses on the information coming from the comparison between 
UV features detected by {\it Swift}-{\tt UVOT}, tracing recent star formation, 
and the galaxy optical structure mapping older stellar populations.  We performed a surface photometric study 
 of  these early-type galaxies, observed with  {\it Swift}-{\tt UVOT}  UV filters,
W2 2030\AA\ $\lambda_0$, M2 2231\AA\ $\lambda_0$, 
W1 2634\AA\ $\lambda_0$, and UBV  bands. 
BVRI photometry from other sources in the literature is also used. Our integrated magnitude
measurements have been analyzed and compared  with  corresponding values in the literature. 
We characterize the overall galaxy structure best fitting the UV and optical luminosity 
profiles using a single S\'ersic law.}
{NGC~1366, NGC 1426, NGC 3818, NGC 3962 and NGC 7192 
show featureless luminosity profiles. Excluding NGC~1366 which has a 
clear edge-on disk  ($n\approx1-2$), and NGC 3818, the remaining three
have S\'ersic's indices $n\approx3-4$ in optical and a lower index in the UV. 
Bright ring/arm-like structures are revealed by UV images and luminosity profiles of  
NGC 1415, NGC~1533, NGC 1543, NGC~2685, NGC~2974 and IC~2006. 
The ring/arm-like structures are different from galaxy to galaxy.
 S\'ersic indices of UV profiles for those galaxies are in the range $n=1.5-3$
both in S0s and in galaxies classified as "bona fide"  ellipticals, such as NGC~2974 and IC~2006. 
We notice that in our sample optical S\'ersic indices are usually
 larger  than in the UV ones. ($M2$-V)  color profiles are
 bluer in ring/arm-like structures  with respect to the galaxy body.}
{The lower values of Sersic's indices in the UV bands with respect to optical ones, 
suggesting the presence of a disk, point out  that the role of the dissipation cannot be 
neglected  in recent evolutionary phases of  these early-type galaxies.} 


\maketitle

\section{Introduction}

The combination of the Far-UV view provided by {\it GAlaxy Evolution EXplorer}
 ({\it GALEX} hereafter) \citep{Martin05} 
with the Sloan Digital Sky Survey (SDSS hereafter) \citep{Stoughton02} 
has greatly undermined the classical 
view of early-type galaxies (Es+S0s=ETGs hereafter) as passively evolving galaxies. 
Statistical studies  \citep[e.g.][]{Kaviraj07,Schawinski07} found that about 30\% of
massive ETGs show recent/ongoing star formation, with the largest incidence in galaxies 
located in low density environments. Only a small percentage of cluster ETGs, $\sim$5/9\%,
shows the classical UV up-turn due to evolved, rising-branch stars 
whereas $\sim$27/43\% shows recent star formation \citep[][respectively]{Yi11,Hernandez14}.

%
\begin{table*}
\caption{General properties of   the sample}
\begin{tabular}{llllllllll}
\hline
 Galaxy  &  Morpho. & Type   & M$_{K}$  &  $\epsilon$ & PA              &  D         &  M(HI) &   Group & $\rho_{xyz}$ \\
  ident.   &  RSA           &              &                    &                          &   [deg]     & [Mpc]  &  10$^9$M$_\odot$  &  name &  [gal Mpc$^{-3}$]\\
\hline
NGC 1366 & S0$_1$(8) & -2.3$\pm$0.7  &-22.59  & 0.54    &      3.8      &   21.1$\pm$2.1 & $<1.0$ & Fornax  Eridanus Cloud & 0.16\\
NGC 1415 & Sa/SBa late &    0.5 $\pm$1.3 & -23.47         & 0.59    &    141.2   & 22.7$\pm$1.5   & 0.9  &       "    & 0.80    \\
NGC 1426 &   E4    & -4.8$\pm$0.5 &-23.22    & 0.32   & 112.5  & 24.1$\pm$2.4 & \dots  &       "    & 0.66\\ 
NGC 1533 & SB0$_2$(2)/SBa & -2.5$\pm$0.6 &-23.97 & 0.37 & 148.9 & 21.4$\pm$2.1 & $7.4^a$ & Dorado cloud & 0.89\\
NGC 1543 &  RSB0$_{2/3}$(0)/a&  -2.0 $\pm$0.8&-24.04   &  0.77   & 93.8    &20.0$\pm$2.0 & 0.8 & " & 0.95  \\
NGC 2685 & S0$_3$(7) pec &  -1.0$\pm$0.8 &-22.64    & 0.46    & 38.0 &  16.0 & 3.0$^b$ & Ursa Major South Spur & 0.13\\
 NGC 2974 &  E4 & -4.2$\pm$1.2&-25.42 & 0.38   &  44.0 &   21.5$\pm$2.2 & 0.7$^c$ & Lynx Cloud & 0.26\\
NGC 3818 &  E5 &-4.6$\pm$0.8 &-23.94    & 0.37   &  95.9  & 36.3$\pm$3.6 & \dots & Crater Cloud & 0.20 \\
NGC 3962 & E1 & -4.8$\pm$0.4  & -25.07    & 0.28   & 10.0 & 35.3$\pm$3.5 & 2.8$^d$ & " & 0.32 \\ 
NGC 7192 & S0$_2$(0)& -3.9$\pm$0.7 & -24.37   & 0.05  &  9.9 &  37.8$\pm$3.8 & $<0.7^d$& Pavo Indus Spur & 0.28\\
IC 2006  & E1   & -4.2$\pm$0.9  &-23.04      & 0.17    & 34.8 & 20.2$\pm$2.0 & 0.3 & Fornax  Eridanus Cloud & 0.82  \\
\hline
\end{tabular}
\footnotesize{
The morphology (col. 2) is  from the \citet{RSA} catalog.  The  morphological type, T (col. 3),  the
ellipticity, $\epsilon$ (col. 5), and the position angle, PA (col. 6), are 
derived from {\tt Hyperleda} \citep{Makarov2014}
with the exclusion of NGC 7192 which is taken from the CGS atlas. The 
K$_s$ absolute magnitude (col. 4) is from 2MASS.  The adopted Distances (col. 7) are derived from 
the Extragalactic Distance Database \citep{Tully09}. The HI masses (col. 8) are obtained 
using the distance in col. 7 and fluxes  from {\tt NED} and  the following references: 
$^a$ \citet{Ryan-Weber2003}; $^b$\citet{Jozsa09}; $^c$ \citet{Kim88};
$^d$ \citet{Serra10}. The Group  identification (col. 9) and the local galaxy density, $\rho_{xyz}$ (col. 10), 
i.e. the density of galaxies brighter than M$_B$= $-16$ mag, are from \citet{T88}. }
\label{tab1}
\end{table*}

In  color magnitude diagrams, built with SDSS and {\it GALEX} magnitudes,
galaxies in clusters show a strong morphological segregation: Spirals lie in the
blue cloud while ETGs are mainly located in the red sequence \citep[e.g.][]{Yi05,Wyder07}.
However, an intermediate area, called "Green Valley", exists, irrespective of the
environment. Recently,   \cite{Mazzei14a} provided a description 
of the galaxy photometric and morphological  evolution 
in the $(NUV-r)~vs.~M_r$ color magnitude diagram (CMD) plane
applying a SPH code with chemo-photometric
implementation to ETGs in the LeoII cloud. Major merging 
as well as galaxy-galaxy interaction are the mechanisms required  to match the global properties
of such ETGs. The simulations  show that these galaxies spend between 3 to
5 Gyr, depending on their luminosity, crossing the Green Valley before they
reach the red sequence and that rejuvenation episodes are more frequent 
in the less massive ETGs.\\
\indent
{\it GALEX} revealed that ETGs with signatures of recent accretion episodes, like shell structures, 
may have a "rejuvenated" nucleus in the Far-UV \citep{Rampazzo07,Marino09}. 
This agrees with results from the study of Lick 
spectroscopic indices by \citet{Longhetti00} and \citet{Annibali07}, concerning ages within
r$_e/8$, i.e. in the nuclear region, and, more recently,
from mid-infrared (MIR hereafter) {\tt Spitzer}-IRS spectral analysis 
within $\approx$2-3~$r_e/8$.  This latter highlights,
in a large fraction of ETGs, the role of Polycyclic Aromatic Hydrocarbon features (PAHs)  
as tracers of  episodes of star formation  with ages in the range between 1 and 2.5  Gyr,
 with the exact values depending on the metallicity. This look-back time corresponds to a 
redshift  $z\lesssim 0.2$ \citep{Bressan06,Kaneda08,Vega10,Panuzzo11,Rampazzo13,Nanni13}.\\
\indent
Furthermore, ETGs can show Far-UV bright
extended structures (e.g. rings, arm-like features, tails), sometimes completely distinct
in shape from the optical galaxy body, and often associated to HI emission
\citep{Jeong09,Thilker10,Marino11c,Marino11a,Salim10,Rampazzo11,Salim12}.
 Simulations suggest that such  UV bright ring/arm-like structures can be
either a transient product of a major   merger \citep{Mazzei14a} or  can result
from the accretion of a small, gas rich companion \citep{Mapelli15}. 
Both observations and simulations concur in indicating that 
{\it galaxy-scale rejuvenation processes} occur at least 
in some ETGs, likely fed by either residual or freshly accreted  gas.\\
\indent
This paper, the second  of a series based on  {\it Swift} 
\citep{swift1,swift2,swift3} multi-wavelength
 ({\tt XRT + UVOT}) observations, is dedicated to tracing signatures
of recent evolution in nearby ETGs. The sample includes only the eleven 
galaxies in \citet[][Paper~I hereafter]{Trinchieri15} observed by {\it Swift}.  
Here, we examine the Near-UV (NUV hereafter) galaxy structures 
revealed by {\tt UVOT}.  Galaxies are selected 
because of the nearly passive stage of evolution of their nuclear region 
\citep{Rampazzo13}.   Most of these galaxies, which
 from the spectrum of their nucleus may be considered as templates of
{\it nearly-dead} ETGs,
reveal a manifold of bright and peculiar structures in our NUV {\it  Swift}-{\tt UVOT} 
images (Figure 7 of  Paper~I). 
{\it What does  NUV tell us about their story?
Is there a common evolutionary framework for these 
apparently different galaxies?}
The comparison between the NUV,  tracing recent episodes of star formation,  
and  the optical spatial structures,  mapping older stellar populations, 
might provide indications about the formation mechanisms of the 
NUV-bright structures  linking recent and past  ETGs evolutionary scenarios.
The study of the galaxy structures 
has recently developed through the multi-wavelength comparison of the S\'ersic
indices derived from the analysis of the luminosity profiles.
\citet{Vulcani14} fitted with a single S\'ersic law
a set of bands, from  $u$ to H,  of a low redshift galaxy sample taken 
from the {\it Galaxy And Mass Assembly (GAMA)} survey.   They examined both
late-type galaxies and ETGs,   i.e $n<2.5$ and $n>2.5$ in their definition. 
Late-types show an increase of the S\'ersic
index with the wavelength \citep[see also][]{Kennedy16} 
unlike ETGs for which $n$ is almost constant
 \citep[see also][]{LaBarbera2010}. They concluded that the variation
of the S\'ersic index with wavelength is connected with the presence of the disk and likely
connected to a radial variation of stellar populations and/or dust reddening.
We aim to extend to NUV, so far unexplored, such analysis using our
 {\it Swift}-{\tt UVOT} data set.\\
\indent
The paper plan is the following. In Section~2 we recall our 
sample and in Section~\ref{Observations} we described 
observations and the reduction techniques adopted. Difficulties 
in an accurate surface brightness analysis of {\tt UVOT} data
are reviewed as well as the reduction packages adopted.
Results,  presented in Section~\ref{Results}, mainly consider the comparison
between the NUV and the optical spatial light distribution. The discussion
 in the light of the literature is given in Section~\ref{Discussion}. 
 Finally, Section~\ref{Summary} summarizes our results and conclusions.

\section{The sample}
\label{Sample}
Table~\ref{tab1} presents the general characteristics of 
our ETG sample.  The Table does not include NGC 1209, present in
Paper~I, since the galaxy does not have {\it Swift} observations. 
All galaxies are  ETGs according to their classification of col.s 2 and 3. 
A transition case between early and later types is NGC~1415 whose 
classification has, however, a large uncertainty. 
Their absolute $K_s$ magnitude range (col. 4), 
$-25.42 \leq M_{K_s} \leq -22.59 $,
suggests a large interval in the stellar mass and
col. 8 indicates that these ETGs are gas rich on average,
with HI masses of the order of 10$^9$ M$_\odot$. 
Columns 9 and 10 refer to their environmental properties. 
The range of the local densities, $0.13 \leq \rho_{xyz}$ (gal Mpc$^{-3}$) $\leq 0.95$ 
 is large, but in general they are in a relatively poor environment.
Galaxies affiliated to the Fornax cluster and the Eridanus cloud
are peripheral enough that they 
are located in regions with densities typical of groups.  
Adopted distances from the Extragalactic Distance Database \citep{Tully09}, quoted 
in column 7, have uncertainties of about 10\%.

The sample includes only galaxies selected 
because in their nuclear region, 2-3$\times$ r$_e$/8 ,
their MIR spectral classes range from 2 to 0 \citep[][and references therein]{Rampazzo13}. 
MIR classes \citep{Panuzzo11} describe the spectral characteristics
of the  nuclear regions. The class-2 spectra show  atomic and 
molecular emission lines plus PAHs features with anomalous 
inter-band ratios; the class~1 spectra show  emission
lines without PAHs, while class~0 spectra are characterized only by the silicate emission 
at $\lambda\sim$10 $\mu$m  and the $\lambda\sim$18 $\mu$m from evolved stars. 
 \citet{Bressan06} show that class~0 spectra characterize  passively evolving nuclei. 
The sample avoids class~3  spectra. At odds with class ~2,  class~3 spectra show
normal PAHs inter-band ratios, typical of star forming galaxies, like Spirals.
The sample does not include also class~4 spectra. These spectra differ from class~2 ones 
because they are dominated by a hot dust continuum and sometimes show high ionization 
lines, both signatures indicating the presence of an AGN.
Summarizing, the class 2 -1 nuclear spectra of our ETGs are consistent 
with traces of a past star formation  
as well as a possible residual AGN activity \citep[see also][]{Vega10}  while those
of class~0 correspond to  a passively evolving stellar population. 

According to  the galaxy morphology, presented for each band in Figure~7 of  Paper~I,
the sample can be divided into two  sets. 
NGC~1366, NGC~1426, NGC~3818, NGC~3962 and NGC~7192 
do not show remarkable feature both in the optical nor in the NUV bands. 
The remaining six ETGs, namely NGC~1415, NGC~1533, NGC~1543, 
NGC~2685, NGC~2974 and NGC  IC~2006, show bright, sometimes peculiar,
ring/arm-like features  in the NUV filters.

\section{Observations and data reduction}
\label{Observations}
{\tt UVOT} is a 30~cm telescope in the {\it Swift} platform operating
both in imaging and spectroscopy modes \citep{Roming05}. 
We observed our ETGs in imaging
in all six  available filters, $W2$ ($\lambda_0 \ 2030$), $M2$ ($\lambda_0 \ 2231$), 
$W1$ ($\lambda_0 \ 2634$), $U$ ($\lambda_0 \ 3501$),
$B$ ($\lambda_0 \ 4329$), $V$ ($\lambda_0 \ 5402$). Description
of the filters, PSFs (FWHM 2\farcs92 for $W2$, 2\farcs45 for $M2$,
2\farcs37 for $W1$, 2\farcs37 for $U$, 2\farcs19 for $B$, 2\farcs18 for $V$),  and 
calibrations are discussed in \citet{Breeveld10,Breeveld11}. 

{\tt UVOT} data obtained in imaging mode with a 2x2 binning, resulting in 1.004"/pixel, 
were processed using the procedure described in {\tt http:// www.swift.ac.uk/analysis/uvot/}.
We combined all the images taken in the same filter for each galaxy 
in a single image using {\tt UVOTSUM} to improve the S/N and to enhance the visibility 
of NUV features of low surface brightness. 

The final data set therefore contains $W2$, $M2$, $W1$, $U$, $B$, $V$
images for each galaxy observed with {\it Swift}, as we discuss below. 
The final exposure times per image are different since 
we complied with the request of preserving the lifetime of the filter wheel and 
we therefore observed as much as possible in the filter-of-the-day.  
Total exposure times, already given in Table~3 of Paper~I, are reported  in
Table~\ref{tab2} for convenience. 
 
We used the  photometric zero points  provided by \citet{Breeveld11} for converting
{\tt UVOT} count rates to the AB magnitude system \citep{Oke74}: 
{\it zp$_{W2}$} = 19.11$\pm$0.03, {\it zp$_{M2}$} = 18.54$\pm$0.03, 
{\it zp$_{W1}$} = 18.95$\pm$0.03, {\it zp$_U$} = 19.36$\pm$0.02, 
{\it zp$_B$} = 18.98$\pm$0.02 and {\it zp$_V$} = 17.88$\pm$0.01.

{\tt UVOT} is a photon counting instrument and, as such, is subject to 
coincidence loss when the throughput is high, whether due to 
background or source counts,  which may result in an undercounting of the flux. 
This effect is a function of brightness of the source and affects
the linearity of the detector.  
The $U$, $B$ and $V$ filters are the most affected 
although coincidence loss can be present also in NUV filters.
In case of our binning,  \citet[][and reference therein]{Hoversten11} 
calculates that count rates less than 0.028 counts~s$^{-1}$~pixel$^{-1}$ 
are affected by at most 1\% due to coincidence loss.  
We checked  the presence of  coincidence loss on 
 all our  images. As an example we show  the count rates of
NGC 1543 in the NUV and optical bands in Figure~1. 
The NUV filters
$UVW2$, $UVM2$ and $UVW1$ are almost free from coincidence loss
effects. We verified that in $UVW1$, the most affected of the NUV filters,
the region is restricted to a few pixels   
centered on the galaxy nucleus. For instance in the case
of NGC 1415 the level of the 0.028 counts~s$^{-1}$~pixel$^{-1}$ 
in the $UVW1$ filter is exceeded in two pixels at the opposite 
sides of the innermost annulus revealed. 

Coincidence loss effects can be corrected in the case of point sources
\citep{Poole2008, Breeveld10}. For extended sources a correction process has been 
performed for NGC 4449, a 
Magellanic-type irregular galaxy with bright star forming regions, by
\citet{Karczewski13}.  Even though
their whole field is affected, the authors calculate that the statistical 
and systematic uncertainties in their total fluxes amount to $\approx$ 7-9\% 
overall,  for the NUV and the optical bands. Based on these results, we decided
not to apply any correction for coincidence loss to our optical data.

\begin{table}[htbp]
   \centering
   \caption{Total exposure times in the  {\tt UVOT } filters } 
   \begin{tabular}{@{} lrrrrrr @{}} 
     
      \multicolumn{7}{c}{} \\
      \hline
    Galaxy   &   W2 & M2 & W1  & U  &   B  & V \\
       Ident.   & [s]   & [s]  & [s] & [s] & [s] & [s] \\
      \hline \hline
   NGC 1366 & 13483 & 11825 &  3597 & 8627 & 1112 & 1112\\
   NGC 1415 & 23990 & 18595 & 22887 &  14857 & 1607 & 1562    \\
   NGC 1426 & 9924  & 5744  & 11179 &  5765  & 544   & 544 \\
   NGC 1533 & 7412  & 7220  &  5607 &  13198   &  1848   & 1848\\
   NGC 1543 & 17414 & 29743 & 20155 &  15133 & 1922   &  1575 \\
   NGC 2685 & 7838  & 6517  &  6456 &   783  &  783   & 783 \\
   NGC 2974 & 16495 & 6709  &  3548 &  7689  & 1082   &  1082  \\
   NGC 3818 & 10616 & 8497  &  3296 &  15392 & 424  & 233\\
   NGC 3962 & 8487  & 11333 & 10490 &   6699 & 808   & 792\\ 
   NGC 7192 & 6927  & 15313 &  6163 & 6787   & 987 & 987\\
   IC 2006  & 12014 & 6238  &  4992 & 11181 & 1431 & 1338\\   
      \hline
   \end{tabular}
\label{tab2}
\end{table}

\begin{figure*} 
\begin{center}
\includegraphics[width=15cm]{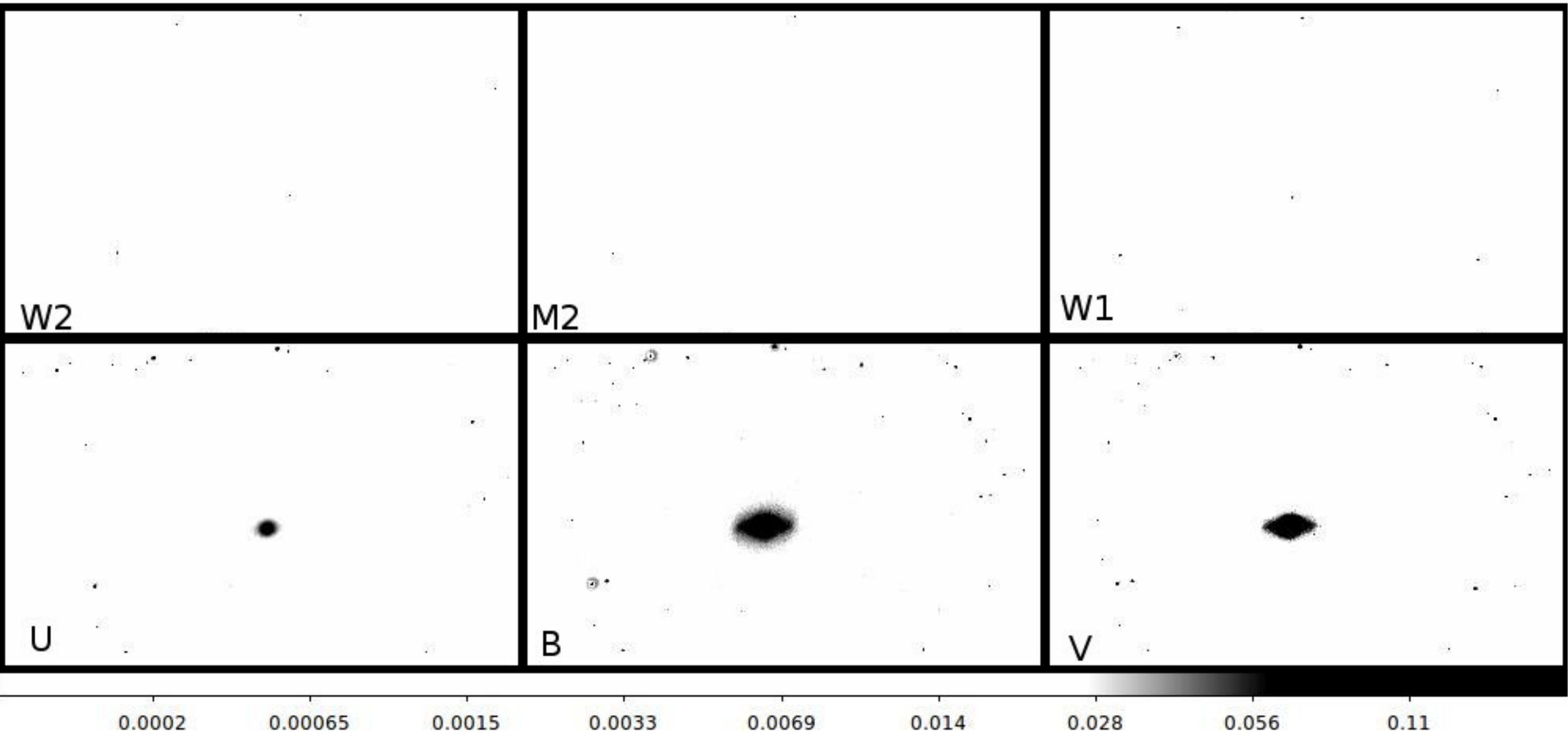}
\caption{UVOT count rate images of NGC 1543 by filter (see composite images in
Figure~\ref{NGC1533-sb} for reference).
 Following \citet{Hoversten11} non-white   areas have a count rate higher than 
0.028 counts s$^{-1}$ pixel$^{-1}$ which indicates that coincidence loss effects 
are larger than 1\% in images  binned 2$\times$2. U, B, V images appear
affected by coincidence loss effects, although at few percent level. At odds, NUV
images are unaffected (see text).}
\end{center}
\label{count-rates}
\end{figure*}

\begin{figure*}
\begin{center}
\includegraphics[width=15cm]{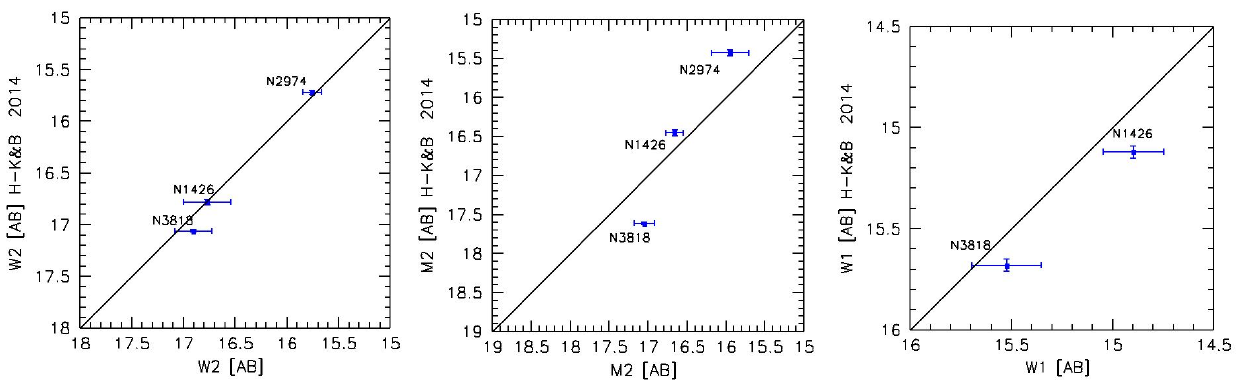}
\caption{Our apparent $W2$, $M2$, and $W1$ integrated magnitudes are compared with those
 derived by \citet{Hodges2014} (galaxies + halo) for the galaxies in common. }
\label{W1M1W2}
\end{center}
\end{figure*}

\begin{figure*}
\begin{center}
\includegraphics[width=16cm]{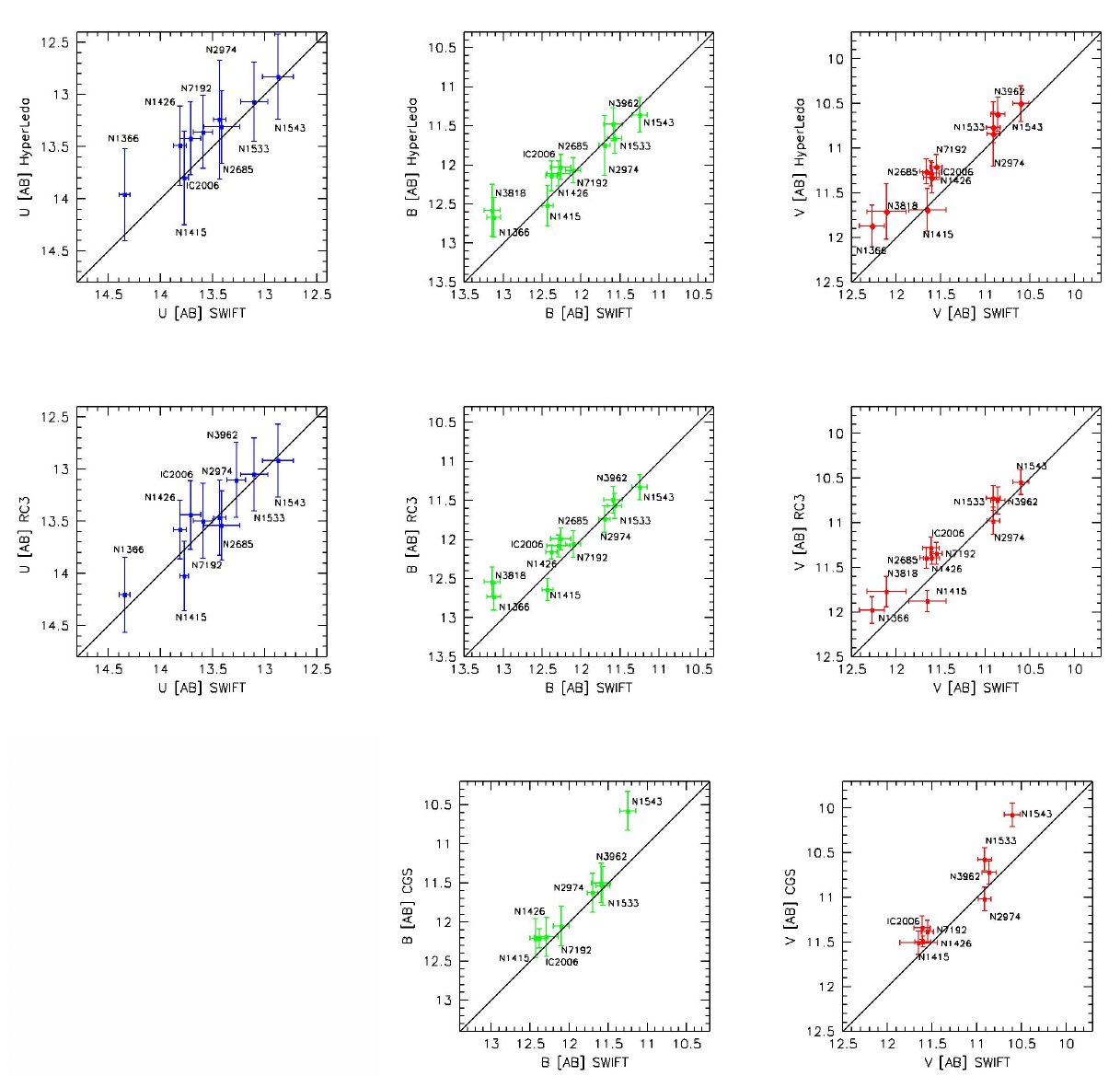}
\caption{Comparison between our apparent integrated  $U$, $B$ and $V$ 
magnitudes and those in the {\tt Hyperleda}  (top panels), the RC3 \citep{deVauc91}
(mid panels) and the CGS \citep{Ho11,Li11}  (bottom panels) catalogues, respectively.  
Magnitudes are in the [AB] system. Error bars of the {\tt Hyperleda} 
and the RC3 $U$ [AB]  magnitudes account for additional $\pm$0.23
magnitude to include the uncertainty in the conversion from 
Vega to [AB] magnitudes. Magnitudes are not corrected 
for Galactic extinction.}
\label{B-Vmag}
\end{center}
\end{figure*}

We considered the presence of instrumental scattered light that,
in the NUV filters, may cover the whole frame,  and of the light scattered 
from stars \citep[e.g.][]{Hodges2014}. This latter may produce 
ghost-images:  for particularly bright stars,
a ring pattern is produced. The most affected frames by both effects are 
the $W2$ and $W1$ filters of NGC 1366 (see \S~\ref{Individual Notes}).  
Our images are  the sum of several (dithered and rotated) frames. 
This sum tends to smooth out large-scale inhomogeneities in the final frame.
 Our targets, however, cover a limited portion of the 17\arcmin$\times$17\arcmin 
 of the {\tt UVOT} field of view. The background, although not homogeneous 
 due to the above factors, can be well evaluated around each object.

Surface photometry has been performed using the 
{\tt ELLIPSE} \citep{Jedr87} fitting routine in the {\tt STSDAS} 
package of {\tt IRAF}, increasing the size of the apertures logarithmically. 
Foreground and background objects 
have been removed substituting them with the surrounding background using the IRAF 
task  {\tt IMEDIT}.  Ghosts of bright stars remain a serious problem for an accurate
surface photometry. Since the problem arises only in the images ($W1$ and $W2$ bands)
of NGC 1366, we have masked them in the same way.
To secure a reliable background measure, we performed the measurement well beyond the galaxy
emission. {\tt ELLIPSE} provides the semi-major axis lengths ($a$) the surface
brightness ($\mu$), the ellipticity ($\epsilon$), the Position Angle (PA) 
and the isophotal shape ($a_4/a$). This  term, labeled $B_4$ in the {\tt ELLIPSE} table, 
provides the deviation from the elliptical shape, parametrized by the fourth cosine 
coefficient of the Fourier expansion of the residuals of the fitting procedure. 
The sign, the absolute value, and the behaviour of $a_4/a$ are indicative of 
the boxyness ($a_4/a<0)$ or diskyness ($a_4/a$>0) 
of the isophotes \citep{Bender89,Capaccioli90,Governato93}. \\
\indent
From the surface brightness profiles, we derived apparent
magnitudes integrating the surface brightness within elliptical isophotes.
 Errors of the NUV and optical magnitudes where estimated by propagating the
statistical errors on the  isophotal intensity provided by {\tt ELLIPSE}.    
Our NUV and optical integrated magnitudes, reported in Table~\ref{tab3},
are not corrected for Galactic extinction. 
NUV luminosity profiles are provided in Table~\ref{tab4}. Optical {\tt UVOT}
luminosity profiles, in the $V$ band for all galaxies and, additionally, in the $U$, $B$ bands 
for NGC~1366, NGC ~2685 and NGC~3818, are also reported in Table~\ref{tab4}. 
These profiles will be used to complete the structural analysis of
our galaxies (see details in Section~\ref{Sersic Analysis}).\\

\section{Results}
\label{Results}

This section focuses on our main results, i.e. the
comparison of our estimated {\it Swift}-{\tt UVOT}  integrated magnitudes
with those available in the literature and the analysis of the NUV surface brightness
photometry of our ETGs .
 
\subsection{Integrated magnitudes}
 In  Figure~\ref{W1M1W2} we show the comparison of our integrated NUV magnitudes 
with those obtained by \citet{Hodges2014}
for  NGC~1426, NGC~2974 and NGC~3818. For consistency we corrected 
our magnitudes for galactic extinction as   \citet{Hodges2014}, i.e.  following \citet{Roming09}.  

Our $W2$ magnitudes are fully consistent with their estimates. Our $W1$ 
magnitudes are about 1\,$\sigma$  brighter. $M2$  values show the largest differences. 
Our data are almost 2\,$\sigma$ fainter than in \citet{Hodges2014}
with the exception of 
NGC 3818 which is brighter. However, this difference  cannot be due to a 
ghost in our M2 images  \citep[their Figure 7]{Trinchieri15} or to
 the effect of scattered light which is negligible in the M2 filter \citep[][]{Hodges2014}.

We compared our integrated $U$,  $B$ and $V$  magnitudes with the RC3
\citep{deVauc91}  and {\tt Hyperleda}\footnote{http://leda.univ-lyon1.fr} catalogues, after
converting their  integrated magnitudes in the AB-system applying the
following correction AB-Vega: $V=-0.01$, $B=-0.13$, $U=0.79$. As an
additional source of comparison we use the integrated magnitudes provided by the
Carnegie-Irvine Galaxy Survey (CGS hereafter, \citet{Ho11,Li11}). 
This B, V, R, I surface photometric study covers a significant fraction  of the 
present sample and has been  performed under good seeing conditions. 
The comparison is shown in Figure~\ref{B-Vmag}. Most of our 
integrated magnitudes are consistent, within errors,  with those of the RC3, {\tt
Hyperleda}, and the CGS.
There are some discrepant cases. Our  $B$ and $V$   integrated magnitudes 
 of NGC 1366 and NGC 3818 are fainter than in the RC3 and the {\tt Hyperleda}
catalogues.  We note that   the NGC~1543 
$B$ and $V$ integrated magnitude measures in the CGS
 are discrepant with both  the RC3 and the {\tt Hyperleda} magnitudes whose  values, in turn,  
agree with our measures. We  estimate that the coincidence loss 
in the central part of the $V$ profile amounts to 0.1-0.2 mag at most 
\citep[see Table A2 of][]{Karczewski13}.\\
\indent
We conclude that our NUV and optical integrated magnitudes in general agree with those
literature. The influence of coincidence loss effect in our optical magnitudes are within 
estimated photometric errors.

\begin{table*}
\centering
\caption{NUV and optical integrated magnitudes in the AB system}
\label{tab3}
   \begin{tabular}{@{} lrrrrrr @{}} 
           \multicolumn{7}{l}{} \\
      \hline
          Galaxy  & W2 & M2 & W1 & U  &   B  & V \\
          \hline\hline
NGC 1366 & 17.55$\pm$0.11 & 17.39$\pm$0.14  & 16.36$\pm$0.07 & 14.34$\pm$0.05 & 13.12$\pm$0.09 & 12.27$\pm$0.14 \\       
NGC 1415 & 15.89$\pm$0.07 & 15.88$\pm$0.13  & 15.00$\pm$0.12 &13.77$\pm$0.04& 12.43$\pm$0.07& 11.65$\pm$0.21 \\  
NGC 1426 &  16.89$\pm$0.23 & 16.79$\pm$0.11&14.99$\pm$0.15   & 13.81$\pm$0.06& 12.38$\pm$0.07 &11.60$\pm$0.09 \\  
NGC 1533 &  15.90$\pm$0.16& 16.07$\pm$0.18  & 14.48$\pm$0.08 & 13.10$\pm$0.13& 11.57$\pm$0.09&10.91$\pm$0.08 \\  
NGC 1543 &   15.36$\pm$0.22& 15.64$\pm$0.17 &14.28$\pm$0.12  & 12.87$\pm$0.15& 11.25$\pm$0.10&10.60$\pm$0.05 \\  
NGC 2685 &  15.53$\pm$0.07& 15.67$\pm$0.13  &14.77$\pm$0.13   & 13.41$\pm$0.17& 12.26$\pm$0.13& 11.66$\pm$0.07 \\
NGC 2974 & 16.17$\pm$0.09& 16.40$\pm$0.24   & 15.00$\pm$0.15   & 13.43$\pm$0.06& 11.70$\pm$0.07& 10.91$\pm$0.07 \\
NGC 3818 & 17.17$\pm$0.18& 17.34$\pm$0.13  & 15.73$\pm$0.17 & 14.10$\pm$0.15& 13.14$\pm$0.10 & 12.11$\pm$0.22 \\  
NGC 3962 & 15.90$\pm$0.15& 16.51$\pm$0.11 & 14.66$\pm$0.11  & 13.27$\pm$0.09& 11.59$\pm$0.12& 10.86$\pm$0.08\\  
NGC 7192 & 16.89$\pm$0.23& 16.79$\pm$0.10 & 15.13$\pm$0.17  & 13.59$\pm$0.09& 12.10$\pm$0.10& 11.55$\pm$0.07 \\  
IC 2066     & 16.22$\pm$0.16&  16.37$\pm$0.20 & 15.20$\pm$0.18 &13.71$\pm$0.10& 12.29$\pm$0.15&11.62$\pm$0.09 \\      
      \hline 
\end{tabular}
\end{table*}

\begin{longtab}
\begin{longtable}{rcccccc}
\caption{\label{tab4} Luminosity profiles in the NUV and optical {\tt UVOT} bands}\\ \hline\hline
     $a$   &   $\mu_{W2}$ & $\mu_{M2}$ & $\mu_{W1}$  & $\mu_{U}$ & $\mu_{B}$ & $\mu_{V}$ \\
     arcsec   & [mag~arcsec$^{-2}$]   & [mag~arcsec$^{-2}$]   & [mag~arcsec$^{-2}$]   & [mag~arcsec$^{-2}$]
     & [mag~arcsec$^{-2}$] & [mag~arcsec$^{-2}$]\\
\endfirsthead
\caption{continued.}\\
\hline\hline
\endhead
\hline
\endfoot
      \hline 
        &    &  &    \\
{\bf NGC 1366}                   &              & &          \\
      &    &  &     \\
1.0     & 22.93$\pm$ 0.01& 23.25$\pm$0.01& 21.51  $\pm$0.01 & 19.69$\pm$0.01 & 18.33$\pm$0.01 & 17.44$\pm$0.01 \\
1.1    & 22.95$\pm$ 0.01&  23.27$\pm$0.01& 21.54  $\pm$0.01 &19.72$\pm$0.01  & 18.37$\pm$0.01 & 17.49$\pm$0.01\\
1.3    & 22.99$\pm$ 0.01& 23.30$\pm$0.01&  21.58  $\pm$0.01 & 19.76$\pm$0.01  & 18.41$\pm$0.01& 17.53$\pm$0.01\\
1.4    & 23.04$\pm$ 0.01&  23.34$\pm$0.01& 21.62  $\pm$0.01 & 19.81$\pm$0.01  & 18.48$\pm$0.01& 17.60$\pm$0.01\\
1.6    & 23.10$\pm$ 0.01& 23.39$\pm$0.01& 21.66 $\pm$0.01   & 19.87$\pm$0.01  & 18.56$\pm$0.01 & 17.69$\pm$0.01\\
1.8    & 23.16$\pm$ 0.01& 23.43$\pm$0.01&21.73  $\pm$0.01   & 19.95$\pm$0.01  & 18.65$\pm$0.01 & 17.78$\pm$0.01\\
2.0   & 23.23$\pm$0.01& 23.49$\pm$0.01& 21.82$\pm$0.01      &  20.05$\pm$0.01 & 18.76$\pm$0.01 & 17.90$\pm$0.01\\
2.2   & 23.34$\pm$0.01& 23.61$\pm$0.01& 21.97$\pm$0.01      &  20.15$\pm$0.01 & 18.90$\pm$0.01 & 18.03$\pm$0.01\\
2.5   & 23.46$\pm$0.01& 23.74$\pm$0.01& 22.12$\pm$0.01      &  20.29$\pm$0.01 & 19.04$\pm$0.01 & 18.17$\pm$0.01\\
2.8   & 23.58$\pm$0.01& 23.86$\pm$0.01& 22.27$\pm$0.01      & 20.44$\pm$0.01  & 19.22$\pm$0.01 & 18.34$\pm$0.01\\
3.2   & 23.71$\pm$0.01& 23.99$\pm$0.01& 22.42$\pm$0.01      & 20.59$\pm$0.01 & 19.41$\pm$0.01 & 18.54$\pm$0.01\\
3.5   & 23.84$\pm$0.01& 24.11$\pm$0.01& 22.57$\pm$0.01      & 20.78 $\pm$0.01 & 19.62$\pm$0.01 & 18.74$\pm$0.01\\
4.0   & 23.99$\pm$0.01& 24.23$\pm$0.01& 22.72$\pm$0.01      & 20.95$\pm$0.01  & 19.83$\pm$0.01& 18.94$\pm$0.01\\
4.5   & 24.16$\pm$0.01& 24.36$\pm$0.01& 22.88$\pm$0.01      & 21.14$\pm$0.01  & 20.05$\pm$0.01 & 19.13$\pm$0.01\\
5.0   & 24.32$\pm$0.01& 24.48$\pm$0.01& 23.03$\pm$0.01      &21.32$\pm$0.01   & 20.23$\pm$0.01 & 19.31$\pm$0.01\\
5.6   & 24.49$\pm$0.01& 24.61$\pm$0.01& 23.18$\pm$0.01      &21.50$\pm$0.01  & 20.38$\pm$0.01  & 19.48$\pm$0.01\\
6.3   & 24.65$\pm$0.01& 24.74$\pm$0.01& 23.34$\pm$0.01      &21.65$\pm$0.01  & 20.50$\pm$0.01 & 19.64$\pm$0.01\\
7.1   & 24.79$\pm$0.01& 24.86$\pm$0.01& 23.49$\pm$0.01      &21.80$\pm$0.01  & 20.63$\pm$0.01 & 19.76$\pm$0.01\\
7.9   & 24.94$\pm$0.01& 24.97$\pm$0.01& 23.64$\pm$0.01      & 21.90$\pm$0.01 & 20.72$\pm$0.01 &  19.83$\pm$0.01\\
8.9   & 25.09 $\pm$0.01& 25.06$\pm$0.01& 23.79$\pm$0.01     & 22.00$\pm$0.01 & 20.78$\pm$0.01 & 19.91$\pm$0.01\\
10.0 &   25.23$\pm$0.01& 25.16$\pm$0.02& 23.91$\pm$0.01    & 22.10$\pm$0.01 & 20.86$\pm$0.01 & 20.02$\pm$0.01\\
11.2 &   25.34$\pm$0.01& 25.30$\pm$0.02& 24.01$\pm$0.01    & 22.19$\pm$0.01 & 20.93$\pm$0.01 & 20.13$\pm$0.01\\
12.6 &   25.44$\pm$0.01& 25.50$\pm$0.02& 24.14$\pm$0.02    & 22.31$\pm$0.01 & 20.98$\pm$0.01 &20.22$\pm$0.01\\
14.1 &   25.60$\pm$0.02& 25.69$\pm$0.02& 24.30$\pm$0.02    & 22.46$\pm$0.01 & 21.10$\pm$0.01 & 20.33$\pm$0.01\\ 
15.8 &   25.81$\pm$0.02& 25.75$\pm$0.02& 24.48$\pm$0.02    &22.63$\pm$0.01  & 21.23$\pm$0.01 & 20.49$\pm$0.01\\ 
17.8 &   25.99$\pm$0.02& 25.88$\pm$0.02& 24.73$\pm$0.02    & 22.84$\pm$0.01 & 21.41$\pm$0.01 & 20.68$\pm$0.01 \\ 
19.9 &   26.20$\pm$0.02& 26.22$\pm$0.02& 25.01$\pm$0.03    & 23.05$\pm$0.01 & 21.61$\pm$0.01 & 20.89$\pm$0.01\\ 
22.4 &   26.38$\pm$0.02& 26.34$\pm$0.03& 25.25$\pm$0.03    & 23.27$\pm$0.01 & 21.87$\pm$0.01 & 21.10$\pm$0.01\\ 
25.1 &   26.65$\pm$0.03& 26.60$\pm$0.03& 25.45$\pm$0.03    &23.51$\pm$0.01 & 22.12$\pm$0.01  & 21.38$\pm$0.01\\ 
28.2 &   26.84$\pm$0.03& 26.80$\pm$0.04& 25.76$\pm$0.05    & 23.78$\pm$0.02 & 22.43$\pm$0.02 & 21.65$\pm$0.01\\ 
31.6 &   27.21$\pm$0.04& 26.99$\pm$0.04& 26.17$\pm$0.06    & 24.08$\pm$0.02 & 22.72$\pm$0.02 & 21.97$\pm$0.02\\ 
35.5 &   27.49$\pm$0.04& 27.29$\pm$0.05& 26.55$\pm$0.07    & 24.42$\pm$0.02 & 23.14$\pm$0.02 & 22.34$\pm$0.02\\ 
39.8 &   27.97$\pm$0.06& 27.52$\pm$0.06& 27.02$\pm$0.11    &24.79$\pm$0.03  & 23.52$\pm$0.02 & 22.73$\pm$0.03\\ 
44.7 &   28.39$\pm$0.09& 27.97$\pm$0.08& 27.74$\pm$0.22    & 25.19$\pm$0.04 & 23.99$\pm$0.03 & 23.09$\pm$0.03\\ 
50.1 &   28.83$\pm$0.14& 28.43$\pm$0.12&  \dots                      & 25.62$\pm$0.05 & 24.94$\pm$0.07& 23.58$\pm$0.04\\ 
56.2 & \dots                     & \dots                 &   \dots                     & 26.26$\pm$0.08 & \dots                  & 24.21$\pm$0.01\\
63.1 & \dots                     & \dots                 & \dots                       & 26.99$\pm$0.16 & \dots                  & 25.18$\pm$0.15\\
            &    &  &   \\
            \hline 
            &    &  &     \\
{\bf NGC 1415}                         &               &  &      \\
            &    &  &     \\
1.0 &   23.38$\pm$0.01 &  23.35$\pm$0.01 &  21.79$\pm$0.01 &    &    & 18.19$\pm$0.02\\
1.1 &    23.38$\pm$0.01&  23.33$\pm$0.01 &  21.79$\pm$0.01 &    &    & 18.21$\pm$0.02\\
1.3 &  23.38$\pm$0.01 &  23.32$\pm$0.01 &  21.79$\pm$0.01  &    &    &  18.23$\pm$0.02\\
1.4 &  23.38$\pm$0.01 &  23.29$\pm$0.01&  21.80$\pm$0.01   &    &    &   18.26$\pm$0.02\\
1.6 &  23.38$\pm$0.01 &  23.26$\pm$0.01&  21.80$\pm$0.01   &    &    &   18.31$\pm$0.02\\
1.8 &  23.37$\pm$0.01 &  23.24$\pm$0.01 & 21.80$\pm$0.01   &     &    & 18.36$\pm$0.02\\
2.0    &    23.37$\pm$0.02   &   23.20$\pm$0.08    &  21.81$\pm$0.05 &    &     & 18.40$\pm$0.02 \\
2.2    &    23.35$\pm$0.02   &   23.16$\pm$0.09    &   21.87$\pm$0.06 &    &    & 18.47$\pm$0.02\\
2.5    &    23.35$\pm$0.03   &   23.10$\pm$0.11    &   21.91$\pm$0.07 &    &    &  18.53$\pm$0.02\\
2.8    &    23.33$\pm$0.03   &   23.00$\pm$0.13    &   21.99$\pm$0.06 &    &    &   18.61$\pm$0.02\\
3.2    &    23.31$\pm$0.03  &   22.97 $\pm$0.13    &  22.07$\pm$0.06  &    &     &   18.69$\pm$0.02\\
3.5    &   23.30$\pm$0.03  &   22.89 $\pm$0.14    &  22.19$\pm$0.04   &    &     &    18.80$\pm$0.02\\
4.0    &  23.26$\pm$0.03 &   22.81 $\pm$0.14    &  22.31$\pm$0.03     &    &     &     18.90$\pm$0.02 \\
4.5    &    23.24$\pm$0.03  &   22.71 $\pm$0.12    &  22.45$\pm$0.02  &    &     &     19.02$\pm$0.02\\
5.0    &    23.22$\pm$0.03 &   22.70 $\pm$0.11    &  22.61$\pm$0.02   &    &     &   19.15$\pm$0.02\\
5.6    &   23.19$\pm$0.03 &   22.73 $\pm$0.08      &  22.75$\pm$0.04  &    &     &   19.3$\pm$0.02\\
6.3    &   23.18$\pm$0.04  &   22.78 $\pm$0.06     &  22.83$\pm$0.05  &    &     &    19.48$\pm$0.02\\
7.1    &   23.17$\pm$0.05   &   22.88$\pm$0.04    &  22.84$\pm$0.06   &    &     &    19.70$\pm$0.02\\
7.9    &   23.18$\pm$0.07   &   23.06$\pm$0.04    &  22.86$\pm$0.06   &    &     & 19.89$\pm$0.02\\
8.9    &   23.20$\pm$0.08 &   23.35$\pm$0.04    &  22.80$\pm$0.06      &   &     &  20.04$\pm$0.02\\
10.0  &   23.24$\pm$0.10   & 23.63$\pm$0.03     &22.85$\pm$0.07      &    &     &  20.21$\pm$0.02\\
11.2  &   23.30$\pm$0.12  & 23.95$\pm$0.03    &   23.02$\pm$0.07      &   &     &  20.37$\pm$0.02\\
12.6 &   23.38$\pm$0.13  &  24.29$\pm$0.04    &   23.37$\pm$0.06      &   &     &  20.54$\pm$0.01\\
14.1 &   23.50$\pm$0.14  & 24.84$\pm$0.04    &   23.89$\pm$0.05       &    &    & 20.67$\pm$0.01\\
15.8 &   23.66$\pm$0.15  &25.43$\pm$0.03     &  24.45$\pm$0.03        &     &   & 20.80$\pm$0.01\\
17.8 &   23.88$\pm$0.15  &  25.55$\pm$0.03 &   24.74$\pm$0.02         &     &   & 20.80$\pm$0.01\\
19.9 &   24.18$\pm$0.14 & 25.47$\pm$0.07 &   24.76$\pm$0.04           &     &   & 21.00$\pm$0.02\\
22.4 &   24.50$\pm$0.1  & 25.87$\pm$0.04  &  24.91$\pm$0.03            &     &   & 21.1$\pm$0.01\\
25.1 &   24.85$\pm$0.11 &25.93$\pm$0.03 &   25.01$\pm$0.02            &     &    & 21.16$\pm$0.02\\
28.2 &   25.13$\pm$0.09 & 26.13$\pm$0.04&   25.17$\pm$0.02            &     &    & 21.24$\pm$0.02\\
31.6 &  25.31$\pm$0.07&  26.27$\pm$0.07 &   25.28$\pm$0.04            &     &    & 21.36$\pm$0.02 \\
35.5 &  25.37$\pm$0.05 &  26.10$\pm$0.09 &   25.29$\pm$0.07           &     &    & 21.40$\pm$0.02\\
39.8  & 25.40$\pm$0.05 &  26.01$\pm$0.13 &   25.36$\pm$0.10           &     &    & 21.45$\pm$0.02\\
44.7  &  25.35$\pm$0.06 &  26.67$\pm$0.12 &   25.89$\pm$0.07          &     &    & 21.75$\pm$0.02 \\
50.1  &  26.17$\pm$0.03 & 27.44 $\pm$0.08 &   26.35$\pm$0.07          &     &    & 22.15$\pm$0.02\\
56.2  &  27.28$\pm$0.03  & 27.96 $\pm$0.07 &   26.75$\pm$0.06         &     &    & 22.66$\pm$0.03\\
63.1  &  27.86$\pm$0.05 & 28.43 $\pm$0.08 &   27.15$\pm$0.05          &     &    & 23.22$\pm$0.04\\
70.8  &  28.20$\pm$0.06 & 28.80 $\pm$0.01 &   27.43$\pm$0.06          &     &    & 23.43$\pm$0.05\\
79.4  &  28.42$\pm$0.07 & 29.32 $\pm$0.15 &   27.61$\pm$0.08          &     &    & 23.67$\pm$0.07\\
89.1  &  28.78$\pm$0.10 & 29.41 $\pm$0.18 &   27.89$\pm$0.09          &     &    & 24.00$\pm$0.09\\
100.0 &  29.0$\pm$0.11 &  29.27$\pm$0.15 &   28.03$\pm$0.11            &     &    & 24.09$\pm$0.09\\
112.2 & 29.20$\pm$0.14 &  29.51$\pm$0.18 &   28.22$\pm$0.13           &     &    & 24.21$\pm$0.10\\
125.9 & 29.37$\pm$0.15 &   29.99 $\pm$0.269&   28.44$\pm$0.15        &     &    & 24.44$\pm$0.13\\
141.3 &  30.20$\pm$0.23 & \dots                        &   28.69$\pm$0.19      &     &    & 24.68$\pm$0.15\\
158.5 & \dots                       & \dots                     & 28.99$\pm$0.23        &     &     & 25.03$\pm$0.21\\
      \hline 
      &    &  &   \\
\end{longtable}
\noindent {Semi-major axis (col. 1), surface brightness and error in $W2$ (col. 2), $M2$ (col. 3) and $W1$ (col. 4). 
The $U$ (col. 5), $B$ (col. 6) and $V$ (col. 7) bands are not corrected for coincidence loss. (continue.)}
\end{longtab}

\begin{longtab}
\begin{longtable}{ccccccr}
\caption{\label{tab5} The NUV structural properties of galaxies}
\\ \hline\hline
&    &  &   &   &       &\\
      band    &   $\langle \epsilon \rangle$ &  $\langle PA \rangle$ &  $\langle a_4/a \rangle$   & $n$  & a$_{(80)}$& Notes  \\
                   &  [1-b/a]                               & [deg]                            & $\times 100$                    &         &  [arcsec]   &\\
 \hline
\endfirsthead
\caption{continued.}\\
\hline\hline
      band    &   $\langle \epsilon \rangle$ &  $\langle PA \rangle$ &  $\langle a_4/a \rangle$   & $n$  & a$_{(80)}$& Notes  \\
                    &  [1-b/a]                               & [deg]                                              & $\times 100$                         &          &  [arcsec]    &\\ 
\hline
        &    &  &   &   &       &\\
         \hline
\endhead
\hline
\endfoot
 &    &  &   &   &       & \\
      &    &  & {\bf NGC 1366}   &   &   &    \\
       &    &  &   &   &   &   \\
W2  &0.40$\pm$0.03    &1.4$\pm$2.4     & -4.3$\pm$14.3  &1.71$\pm$0.01   & 34.7 & (1) \\
M2  & 0.40$\pm$0.09   &2.2$\pm$1.8     &  1.4$\pm$4.4    &1.78$\pm$0.08   & 42.2 & (1)\\
W1 &  0.47$\pm$0.15  & 179.8$\pm$3.7 & 1.8$\pm$3.1     &1.51$\pm$0.05   &  23.5 & (1) \\
              &    &  &   &   &   &    \\
  \hline
            &    &  &   &   &   &    \\
        &    &  & {\bf NGC 1415}  &   &   &     \\
         &    &  &   &   &    &   \\
W2  & 0.60$\pm$0.17    &141.2$\pm$6.3  & -34.9$\pm$60.0  & 2.46$\pm$0.19  &  74.4  &(1) \\
M2 &0.33$\pm$0.12   &143.6$\pm$24.8  &  8.7$\pm$36.1 &1.66$\pm$0.05       &  57.4  & (1)\\
W1   & 0.34$\pm$0.09   & 137.1$\pm$19.5 & 0.5$\pm$18.2   &0.85$\pm$0.05   &  98.2  & (1)\\
               &    &  &   &   &   &    \\
                   \hline
                    &    &  &   &   &    &   \\
                &    &  & {\bf NGC 1426}  &   &  &     \\
                 &    &  &   &   &       &\\
W2  &0.30$\pm$0.05    &103.6 $\pm$19.4 & 0.4$\pm$10.3   &  2.65$\pm$0.13   &61.7 & (1) \\
M2 & 0.30$\pm$0.04    &109.5$\pm$18.2 & -1.0$\pm$9.5   & 2.71$\pm$0.10      & 82.7 & (1) \\
W1& 0.29$\pm$0.06   & 117.7$\pm$34.59 & -0.5$\pm$5.1   & 2.86$\pm$0.04     & 148.5 & (1) \\
                    &    &  &   &   &  &     \\
                           \hline
                            &    &  &   &   &   &    \\
         &    &  & {\bf NGC 1533}    &   &   &    \\
          &    &  &   &   &   &    \\
W2  & 0.22$\pm$0.06 & 160.2$\pm$2.5 & -3.3$\pm$8.0  & 2.76$\pm$0.10        &85.5 & (1)$^a$ \\
M2 & 0.20$\pm$0.07 & 174.6$\pm$12.9 & -1.8$\pm$7.6  &  2.54$\pm$0.06      & 85.5 & (1)$^a$ \\
W1&  0.12$\pm$0.09  & 130.7$\pm$51.7 & 0.2$\pm$4.3  &    3.74$\pm$0.09    & 113.7 & (1)$^a$ \\
                             &    &  &   &   &   &    \\
            \hline
             &    &  &   &   &   &    \\
         &    &  & {\bf NGC 1543}  &   &    &   \\
                  &    &  &   &   &  &     \\
W2  & 0.20$\pm$0.09 & 120.5$\pm$40.8 & 2.8$\pm$10.3  & 2.81$\pm$0.16  &197.3 & (2)\\
M2  &0.15$\pm$0.10 &  130.7$\pm$53.5  & 4.3$\pm$13.8  &  2.82$\pm$0.16 & 187.9 & (2)\\
W1&   0.21$\pm$0.11 & 130.4$\pm$53.2  &  2.3$\pm$9.0 &  2.21$\pm$0.15 & 179.0  &  (2)\\
                &    &  &   &   &    &   \\
                  \hline
                   &    &  &   &   &  &     \\
               &    &  & {\bf NGC 2685}  &    &   &  \\
                &    &  &   &   &   &    \\
W2  &0.47$\pm$0.02 & 35.1$\pm$1.12 & -4.1$\pm$27.1    & 2.50$\pm$0.04  & 102.0 &(1)\\
M2 &0.55$\pm$0.22 & 54.6$\pm$35.6 &-13.7$\pm$21.3    &1.60$\pm$0.04   & 86.1 &(1)\\
W1&  0.44$\pm$0.07  & 39.8$\pm$28.8 &-0.6$\pm$6.6   & 2.49$\pm$0.11  & 87.2 & (1)\\
                      &    &  &   &   &   &    \\
               \hline 
                &    &  &   &   &   &    \\
                    &    &  & {\bf NGC 2974}  &    &    &   \\
                     &    &  &   &   &  &     \\
W2  & 0.38$\pm$0.06   &45.2$\pm$2.5  &-0.3$\pm$7.9    &  3.31$\pm$0.35 &     82.7  & (2)$^a$ \\
M2 &  0.24$\pm$0.06  &40.6$\pm$4.7   & 3.2$\pm$10.4   &  3.23$\pm$0.15 &   78.8  & (2)$^a$ \\
W1&  0.35$\pm$0.04  &46.5$\pm$4.1  & 0.6$\pm$3.9   & 3.43$\pm$0.08     &   82.7 & (2)$^a$ \\
                        &    &  &   &   &   &    \\
                       \hline
                        &    &  &   &   &   &    \\
                      &    &  & {\bf NGC 3818}  &   &    &\\
                       &    &  &   &   &  &     \\
W1  &0.33$\pm$0.06  & 108.0$\pm$34.1 &  1.0$\pm$5.7 &  3.59$\pm$0.25   &58.8 & (1) \\
M2 & 0.23$\pm$0.05   &108.2$\pm$27.3  &2.5$\pm$8.8 &  2.73$\pm$0.18    & 56.0 & (1) \\
W2&  0.32$\pm$0.06  &101.3$\pm$32.9  &1.3$\pm$8.3   &  3.33$\pm$0.20  &  75.0  & (1) \\
                          &    &  &   &   &   &    \\
 \hline
        &    &  &   &   &  &     \\
                      &    &  & {\bf NGC 3962}  &   &    &   \\
                       &    &  &   &   &   &    \\
W2  & 0.24$\pm$0.07 & 15.4$\pm$5.0     &-2.4$\pm$6.0   & 3.28$\pm$0.07    & 122.2 & (1) \\
M2   & 0.22$\pm$0.02   & 8.8$\pm$2.5    & -2.4$\pm$1.0   & 2.61$\pm$0.14   &  75.0 & (1) \\
W1   &  0.18$\pm$0.04  & 4.5$\pm$28.0  & -2.9$\pm$5.6  &  2.52$\pm$0.04   & 105.5 &(1) \\
                             &    &  &   &   &  &     \\
 \hline
        &    &  &   &   &   &    \\
                      &    &  & {\bf NGC 7192}  &   &   &    \\
                       &    &  &   &   &    &   \\
W2  & 0.09$\pm$0.05 & 15.5$\pm$57.8 & 0.6$\pm$6.5     & 2.69$\pm$0.21  & 50.8 & (1) \\
M2 &  0.08$\pm$0.06 & 26.6$\pm$69.5 & 1.1$\pm$9.2     & 3.26$\pm$0.18  & 61.7  & (1) \\
W1& 0.06$\pm$0.03   &  146.8$\pm$80.3 & 0.3$\pm$6.1   &  2.98$\pm$0.06& 100.5  & (1) \\
                             &    &  &   &   &   &    \\
        \hline
         &    &  &   &   &    &   \\
                       &    &  & {\bf IC 2006}  &   &    &   \\
                        &    &  &   &   &   &    \\
W2  & 0.08$\pm$0.03  & 32.2$\pm$29.8    &-0.3$\pm$4.2     &  2.21$\pm$0.10   & 137.6 & (2)\\
M2 & 0.13$\pm$0.05   & 177.4$\pm$60.4  &-2.5$\pm$10.7   &  3.36$\pm$0.10  &  151.4 & (2)\\
W1&  0.07$\pm$0.02  & 58.2$\pm$21.0     & -0.4$\pm$6.4    &  2.11$\pm$0.06   & 113.8 &(2)\\
                              &    &  &   &   &   &    \\
                              \hline \hline                                                                                             
\end{longtable}
\noindent {The ellipticity (col. 2), the Position Angle (col. 3) and the isophotal shape, $a_4/a\times100$, (col. 4) are 
the mean values calculated in the same range of the luminosity profile, 
excluding the central $2\times$FWHM$_{PSF}$. 
The errors correspond to the standard deviation. The S\'ersic index of the single fit
is given in column~5. The semi-major axis including 80\% of the integrated
magnitude is given  in column~6. Notes: The fit on optical and NIR profiles is made on the same 
extension as in the NUV profiles. In particular: (1) The single S\'ersic law fit 
is performed  without masking.  
(1)$^a$ NGC 1533: the S\'ersic fit is performed without masking the NUV profile 
(Figure~\ref{NGC1533-sb}) while in BVRI we mask the region 20\arcsec-90\arcsec\ where the bar and the 
lens are present.  (2) The S\'ersic fit excludes the 
ring in NUV and BVRI profiles as shown in  Figure~\ref{NGC1543-sb} and Figure~\ref{IC2006-sb}.
(2)$^a$ The NGC 2974 fit in the NUV is performed masking the region from 30\arcsec\ to 100\arcsec 
(Figure~\ref{NGC2974-sb}) while in BVRI no masking has 
been applied. See Section~\ref{Individual Notes} for details about $B$, $V$, $R$, $I$ S\'ersic fits.} 
\end{longtab}

\subsection{The NUV-{\it Swift} surface photometry}
\label{UV-view}

\begin{figure*} 
\center
\includegraphics[width=15.5cm]{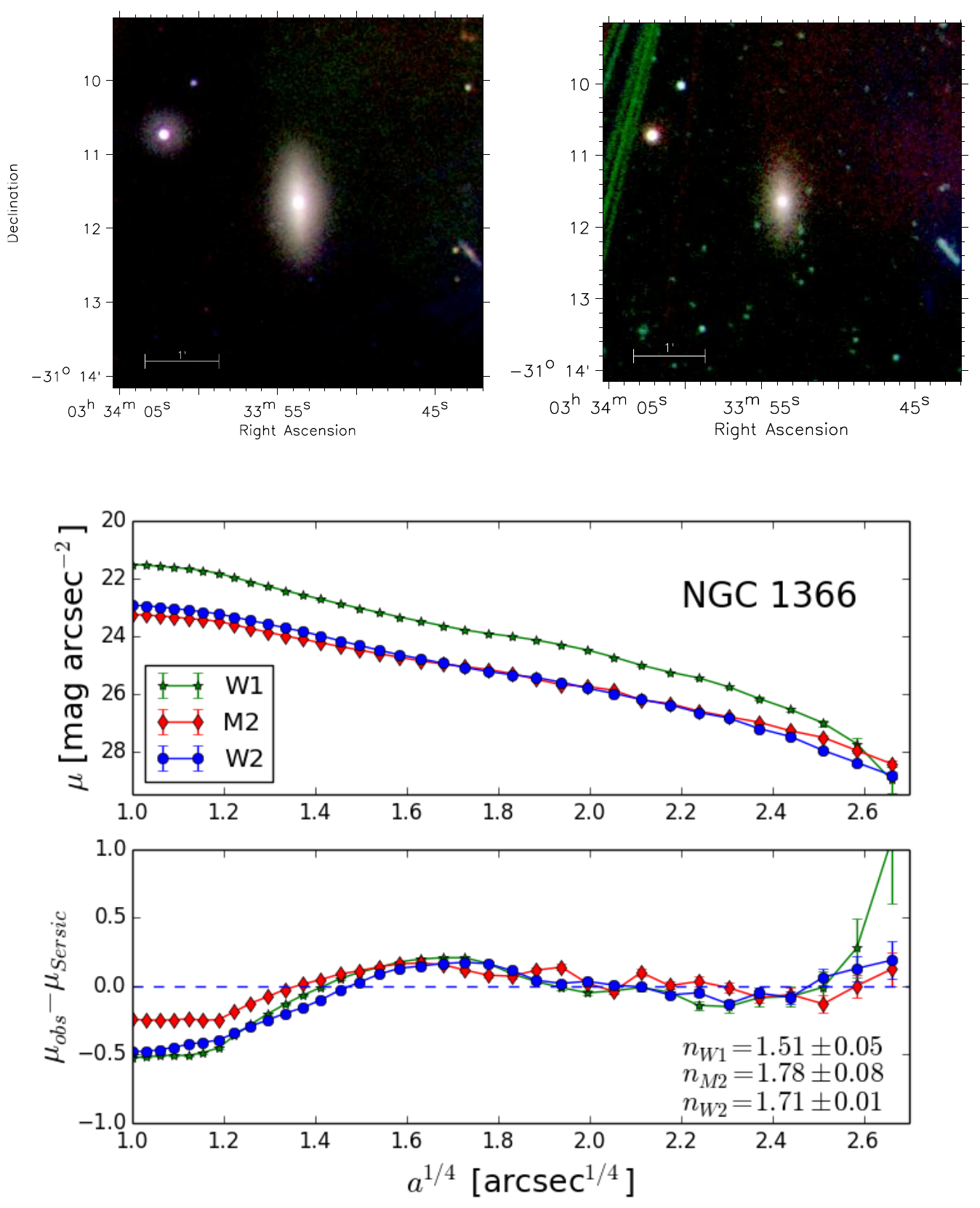}
\caption{NGC~1366. Top panels: Color composite  image   in the U, B, V filters (left: U=blue,
 B=green, V=red) and W1, M2 and W2 filters (right: W2=blue, M2=green,
 W1=red). The field of view is 5\arcmin$\times$5\arcmin, North is on the top, East to 
the left. Middle and bottom panels:
$W1$, $M2$ and $W2$ luminosity profiles and residuals from their 
best fit with a single S\'ersic law accounting for the filter PSF. }
\label{NGC1366-sb}
\end{figure*}

Pictures in the top panels of Figures~\ref{NGC1366-sb} to \ref{IC2006-sb}
compare  the optical  and  the UV morphologies showing color
composite multi-wavelength images of our ETGs. 
Their luminosity profiles in the $W2$, $M2$ and $W1$ bands 
are plotted  in the middle panels of these figures and are  tabulated 
 in Table~\ref{tab4},  truncated when the uncertainty in the surface brightness 
exceeds 0.3 mag~arcsec$^{-2}$.  As for the integrated magnitudes 
they are not corrected for galactic extinction. 

To parametrize the shape of our NUV luminosity profiles we adopt
a S\'ersic r$^{1/n}$ law \citep{Sersic68}, widely used for elliptical and S0 galaxies 
since it is a generalization of the r$^{1/4}$ \citet{deVauc48} law  \citep[see e.g.][]{Caon93}. 
Special cases are $n$=1, the value for an exponential profile,
and $n=0.5$, a Gaussian luminosity profile.
 Galaxies with $n$ values larger than $1$ have a steep luminosity profile in their
nuclear regions and extended outskirts. Values smaller than $1$  indicate a flat nuclear region and 
more sharply truncated outskirts.   From a 2D  luminosity profile decomposition  of about  
200 ellipticals from the SDSS, \citet{Gadotti09}
 measured that the S\'ersic index in $i$-band has a mean value of
 $3.8\pm0.9$, close to $n=4$, the historical paradigm for 
 {\it bona fide} ellipticals \citep{deVauc48}, although with a large scatter.\\
\indent
We best fit a S\'ersic law convolved with a PSF, using a custom {\tt IDL} routine
based on the  {\tt MPFit} package \citep{Markwardt2009},
accounting for errors in the surface photometry.   
The PSF model is a Gaussian of given FWHM and the convolution is computed using 
FFT on oversampled vectors. We use the nominal value of the FWHM of the PSF for
each {\tt UVOT} filter.  However, due to the co-adding, binning and relative rotation
of the frames the FWHM is broadened $\approx$15\% from the nominal value \citep[][]{Breeveld10}.
We verified that the effects of
the small variations of the PSF are well within the error associated to the S\'ersic index. 
The residuals, $\mu - \mu_{Sersic}$, are shown in the bottom panels of 
Figures~\ref{NGC1366-sb} to \ref{IC2006-sb}, the S\'ersic indices 
collected in Table~\ref{tab5} and shown in Figure~\ref{analysis-1}. \\ 
\indent
The structural properties of our galaxies in
the NUV bands are collected in Table~\ref{tab5}.
Columns 2, 3, 4 provide the ellipticity, the position angle (PA)
and the isophotal shape parameter, $a_4/a$, averaged over $a_{80}$ 
(col. 6) which is the semi-major axis including  80\%
of the total galaxy luminosity. 
In this average we exclude the values within $2 \times FWHM_{PSF}$ of the 
galaxy centers to minimize  PSF effects.
The quoted error for each measure is the standard deviation around the mean. 
Figure~\ref{a4M2} shows  the isophotal shape 
parameter  $a_4/a$ as a function of the semi-major axis in the $M2$ band. 
Due to the presence of ring/arm-like structures strong variations are present and 
motivate our choice to provide the average values of $a_4/a$, PA and ellipticity, 
as explained above.  Table~\ref{tab5} also reports  the S\'ersic index
resulting from the best fit of the luminosity profile (col. 5) and the notes about the fit (col. 7).

We show in Figure~\ref{color-M2-V} the ($M2$-V) color profiles
obtained from our {\tt UVOT}-$M2$  and $V$-bands data, i.e. we compare
data sets with a similar PSF, avoiding to use the CGS because of 
the PSF mismatch. We choose the $M2$ band over $W1$ and $W2$ because
is  less affected by scattered light and  ghosts (see \S~2). 

In the next sub-section, we discuss our NUV photometric 
results for each single object in the context of the current literature.

\subsection{S\'ersic index analysis}
\label{Sersic Analysis} 
The study of the S\'ersic index is usually performed on "un-structured" ETGs
like NGC 1366, NGC 1426, NGC 3818, NGC 3962 and NGC~7192.
ETGs with ring/arm-like structures have either knotty, irregular or filamentary 
morphologies as expected for sites hosting young stellar populations. These
features reflect on their luminosity profiles and sometimes 
they become  the dominant characteristics in the NUV luminosity
profile. We are however interested in obtaining  
a description of the  galaxy structure underlying these features  
which may differ when moving from UV to optical bands. For this reason
we fit and compare S\'ersic indices in different bands considering,
case by case, the regions where these features occur (see Section~\ref{Individual Notes}).
 This strategy allows us to follow the variation of the underlying  galaxy 
structure, if any, across different wavelengths. 

 We have decided to use CGS data  \citep{Ho11, Li11} to investigate the 
variation of the S\'ersic index with wavelength for several reasons.  
All galaxies, except NGC~1366, NGC~2685 and NGC~3818 have B, V, R, I photometry 
in  the CGS, taken with good seeing conditions. They report a median FWHM
of the seeing of 1\farcs17,  1\farcs1, 1\farcs01, 0\farcs96 in the B, V,
R, I bands, respectively, better than our {\tt UVOT} $B$ (FWHM=2\farcs18) and $V$
(FWHM=2\farcs19). In few cases the CGS optical luminosity profiles extend further than ours.
Finally,  the R and I filters allow  us to investigate the galaxy photometric structure 
from NUV to NIR.
For NGC~1366, NGC~2685 and NGC~3818, not in the CGS sample, we
use our {\tt UVOT} $U$, $B$, $V$ luminosity profiles to complete the analysis
of the sample (see Table~\ref{tab4}). \\
\indent
To be consistent with our analysis of the UV data, we 
do not follow multi-component decomposition approaches discussed in the literature
 such as fitting the profile with multiple   S\'ersic law profiles by \citet{Huang13} 
 \citep[see also][]{Laurikainen06,Laurikainen10,Laurikainen11}.
 We  performed  the analysis of the CGS profiles with a 
single S\'ersic law fit, convolving the profile with the median
FWHM of the seeing.  This choice allows us to use 
the same criteria adopted for the NUV, i.e. masking the same structures,
 and extending the fit to the same regions (see notes in column 8 of 
Table~\ref{tab5}) for 9 out of 11 galaxies. For the remaining two galaxies, namely NGC 1533
and NGC 2974, in order to map the underlying galaxy structure we mask
the sub-structures present  in optical but not in the NUV profiles,  or viceversa,
as explained in the individual notes of these galaxies. \\
\indent
The synoptic view of the results of the single S\'ersic fit for the different
bands, from NUV to NIR, is shown in Figure~\ref{analysis-1}. This figure
shows that S\'ersic indices vary from filter to filter for a given galaxy and also
vary from galaxy to galaxy. Few of the indexes plotted comply with the
"canonical" value of $n=4$, highlighted with the arrow in the figure. Before
entering in the discussion of these results in the context of the current literature,
in the next section we discuss the NUV photometric results in detail for each galaxy.

\subsection{Notes on individual galaxies}
\label{Individual Notes}

\medskip\noindent
\underbar{NGC 1366}~~~ 
A  bright star is present to the North. Light scattered by the instrument 
and a ghost of the star projects in the North-West region around the galaxy in the $W1$ 
the $W2$ bands.  The irregular background could  perturb our photometric 
study in these two bands.

Both the RC3 \citep{deVauc91} and the RSA
\citep{RSA} classify NGC 1366 as an S0. Recently \citet{Buta15} classified
NGC 1366 as SA(rl)0$^-$ sp in the survey S$^4$G  adding the note that 
a very subtle inner ring-lens (rl) is present \citep[see also][]{Comeron14}.   
Our surface photometry extends out to 50\arcsec\ ($\sim 5$ kpc using the distance in Table~\ref{tab1}), 
i.e. about 2$\times$r$_{eff}$ in NUV.  

The $n$ values (Figure~\ref{NGC1366-sb} bottom panel) are lower than 2
in all the NUV bands, consistent  with the presence of a disk structure. Very
similar values are also measured in the $U$ and $V$ bands while the $B$ band 
gives a larger value of $n=2.55\pm0.03$ 
(Figure~\ref{analysis-1}).  The nuclear part of the galaxy is brighter    
than the S\'ersic model in all  the  NUV bands.

The $a_4/a$ ratio in $M2$ (Figure~\ref{a4M2}) shows a boxy nucleus and 
disky outskirts. 

The ($M2-V$) color profile   of this galaxy (Figure~\ref{color-M2-V}) is bluer
in the outer parts relative to  the inner ones.

\begin{figure*} 
\center
\includegraphics[width=15.5cm]{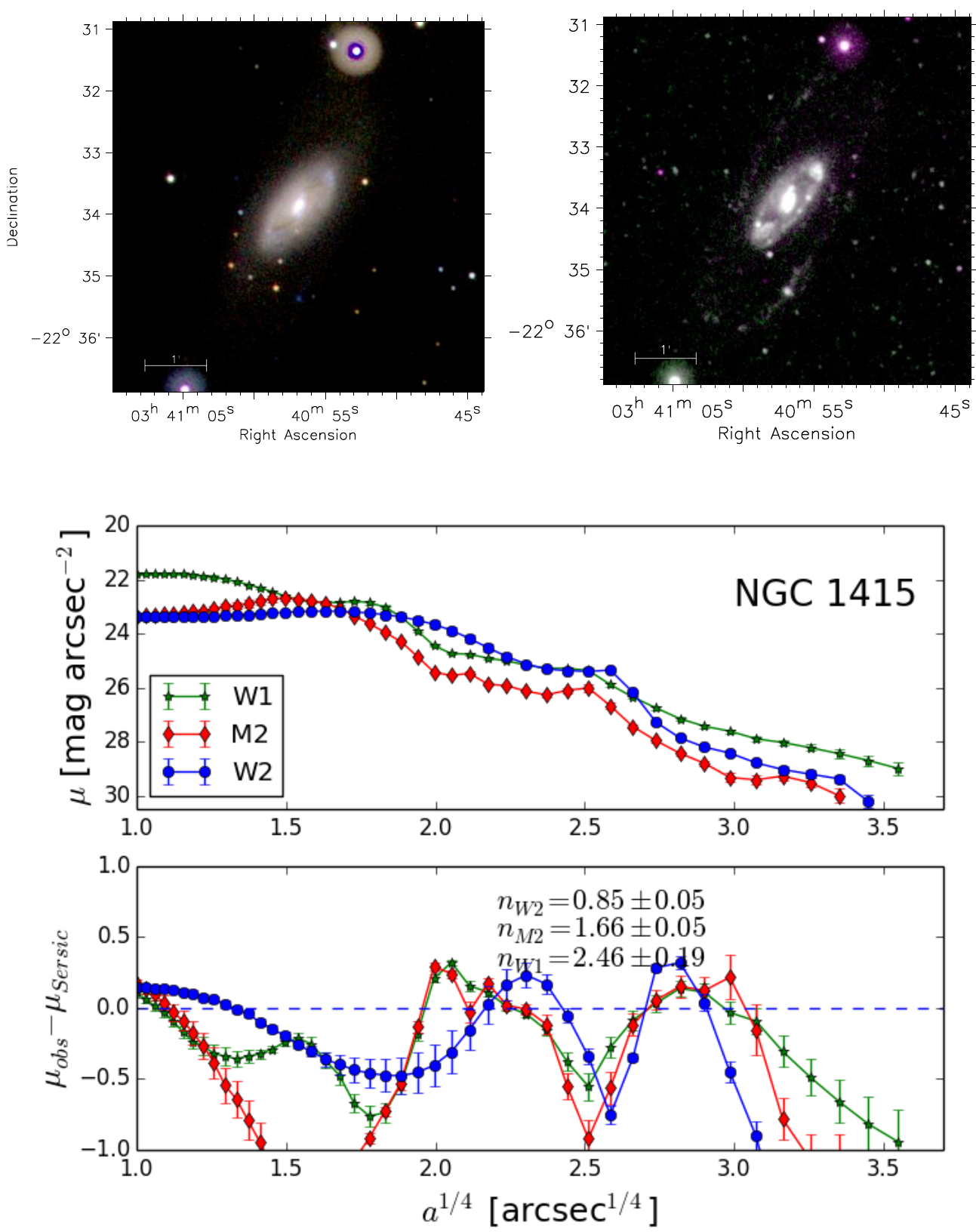}
	\caption{As in Figure~\ref{NGC1366-sb} for NGC~1415. 
	The field of view is 7\arcmin$\times$7\arcmin. The wide rings
around bright sources in the left panel, are an artifact of the {\tt UVOT}
 on-board centroid algorithm in the high count rate regime \citep[see also][and references
  therein]{Hoversten11}.}
\label{NGC1415-sb}
\end{figure*}

\begin{figure*} 
\center
\includegraphics[width=15.5cm]{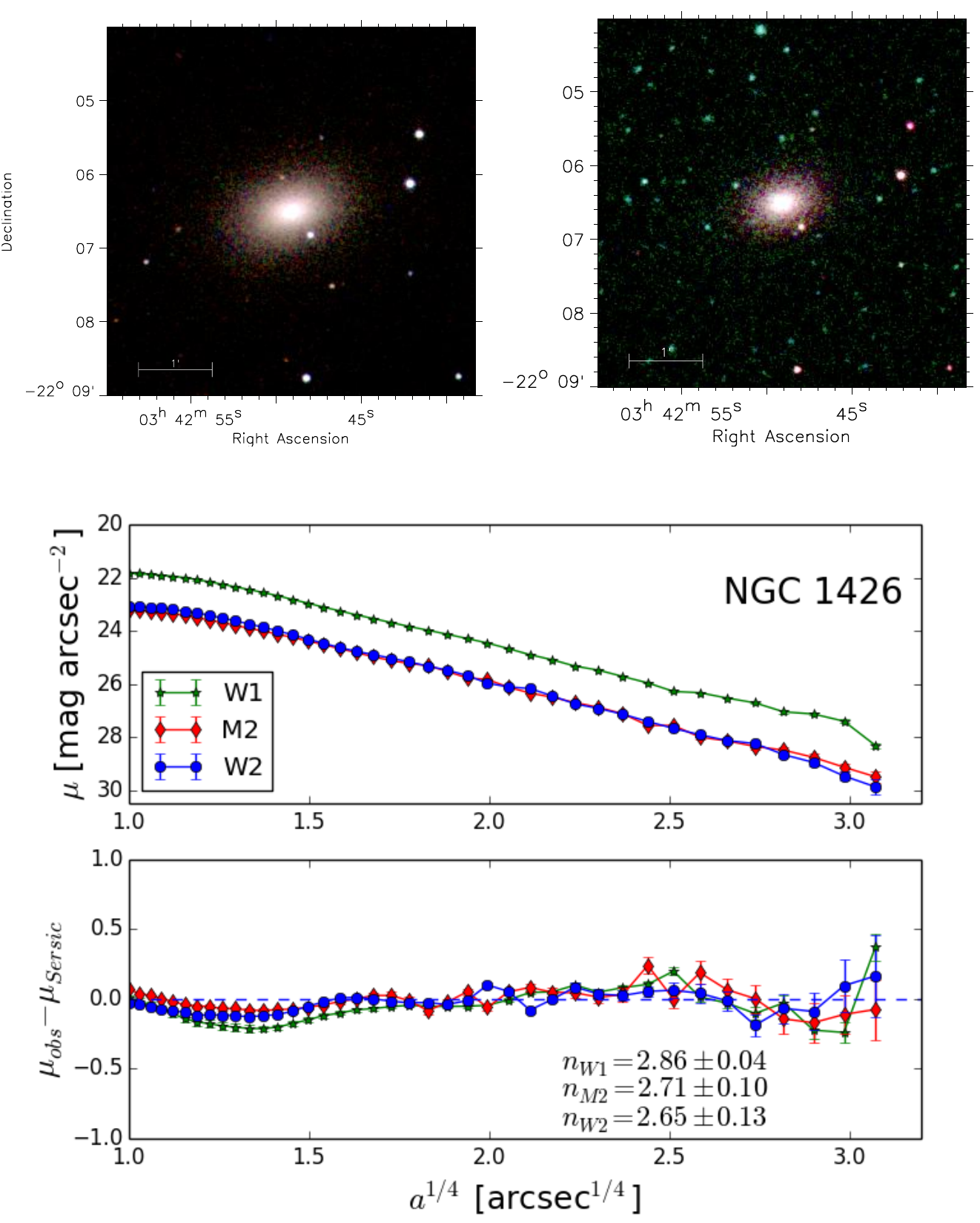}
\caption{As in Figure~\ref{NGC1366-sb} for NGC~1426. The field of view is 
5\arcmin$\times$5\arcmin.}
\label{NGC1426-sb}.
\end{figure*}

\begin{figure*} 
	\center
\includegraphics[width=15.5cm]{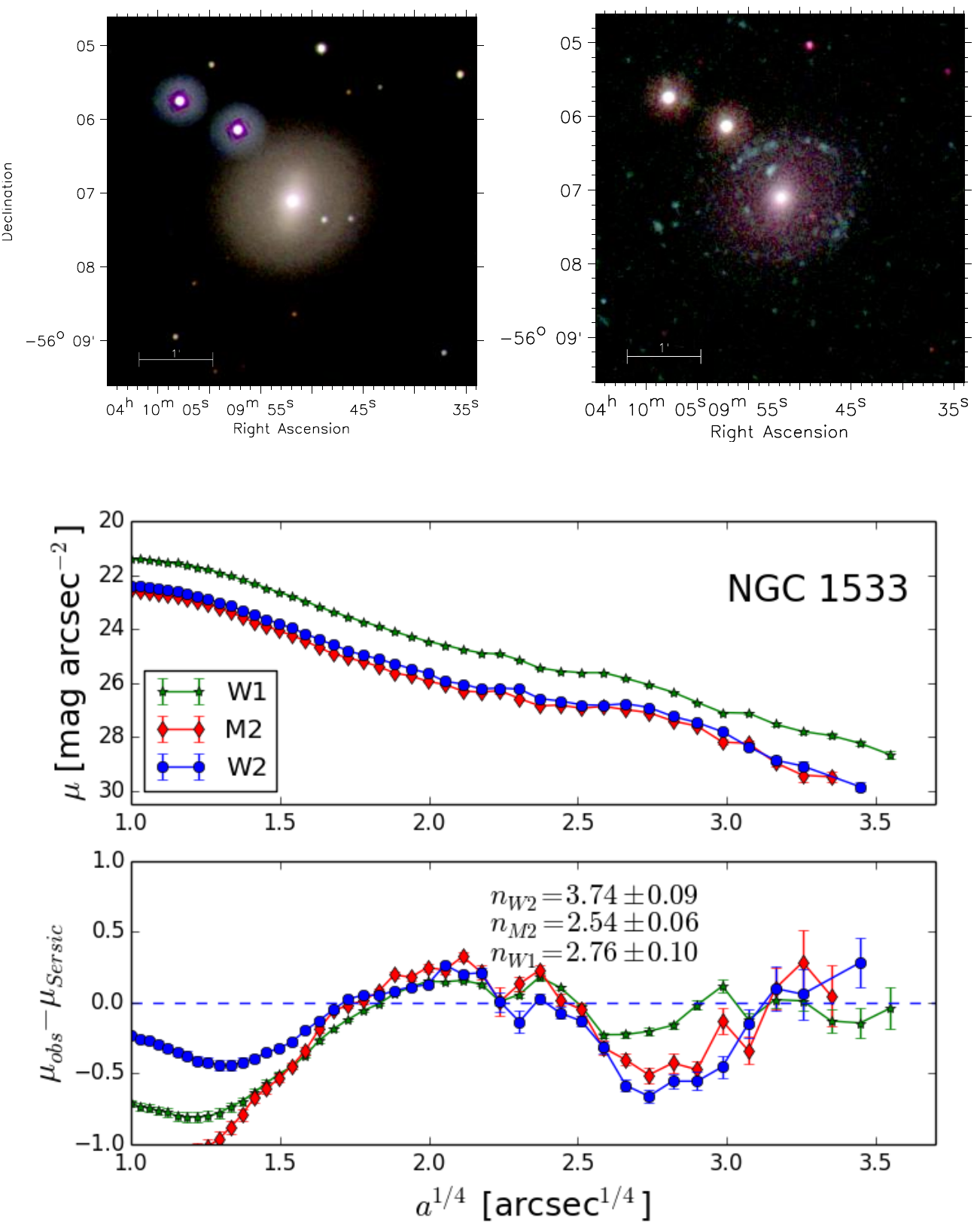}
\caption{As in Figure~\ref{NGC1366-sb} for NGC~1533.The field of view is
 5\arcmin$\times$5\arcmin.}
\label{NGC1533-sb}
\end{figure*}
\begin{figure*} 
	\center
\includegraphics[width=15.5cm]{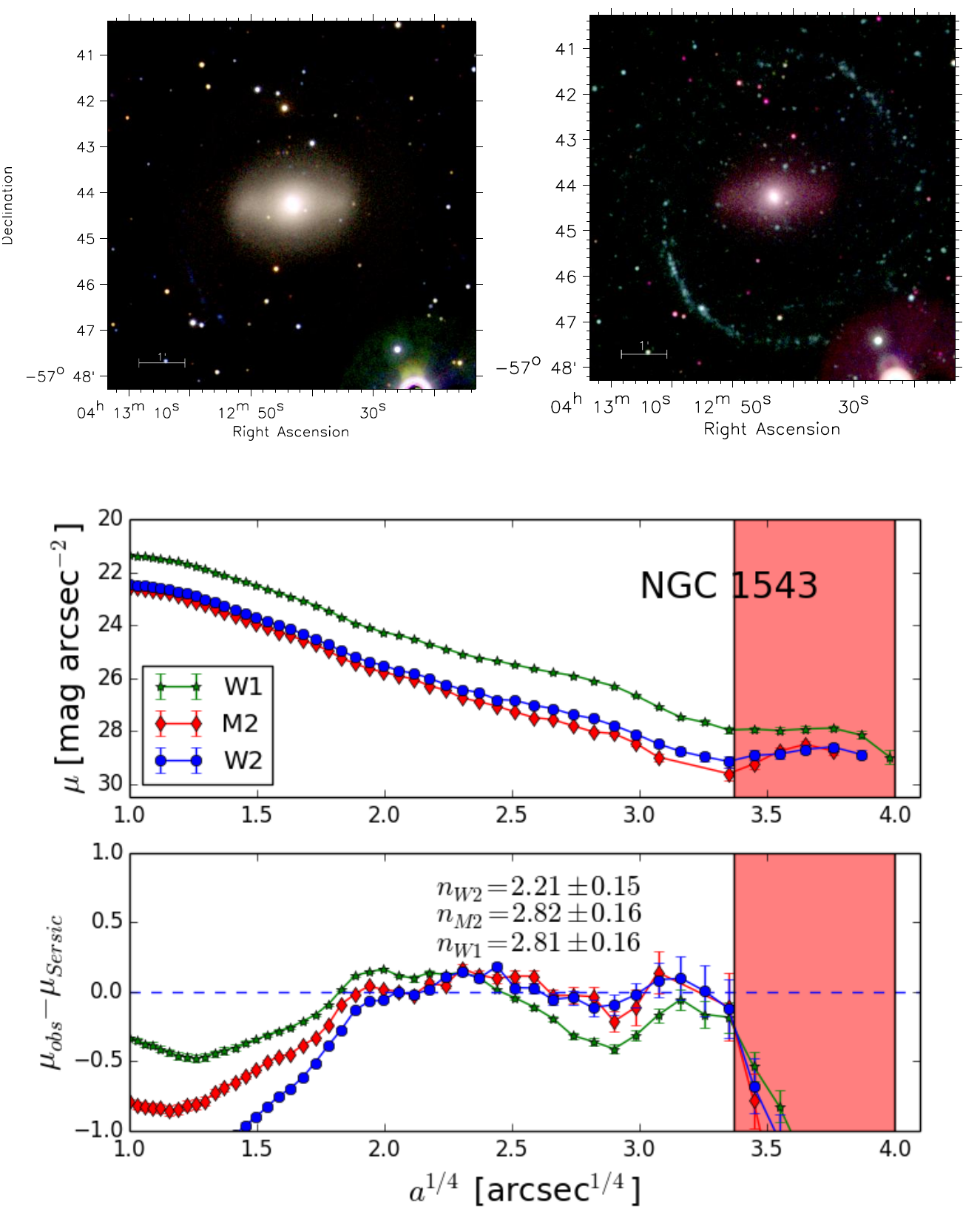}
	\caption{ As in Figure~\ref{NGC1366-sb} for NGC~1543. The field of view is 
	8\arcmin$\times$8\arcmin. The shaded area indicates the masked region in the
	S\'ersic law fit (see text)}
	\label{NGC1543-sb}
\end{figure*}
\begin{figure*} 
	\center
\includegraphics[width=15.5cm]{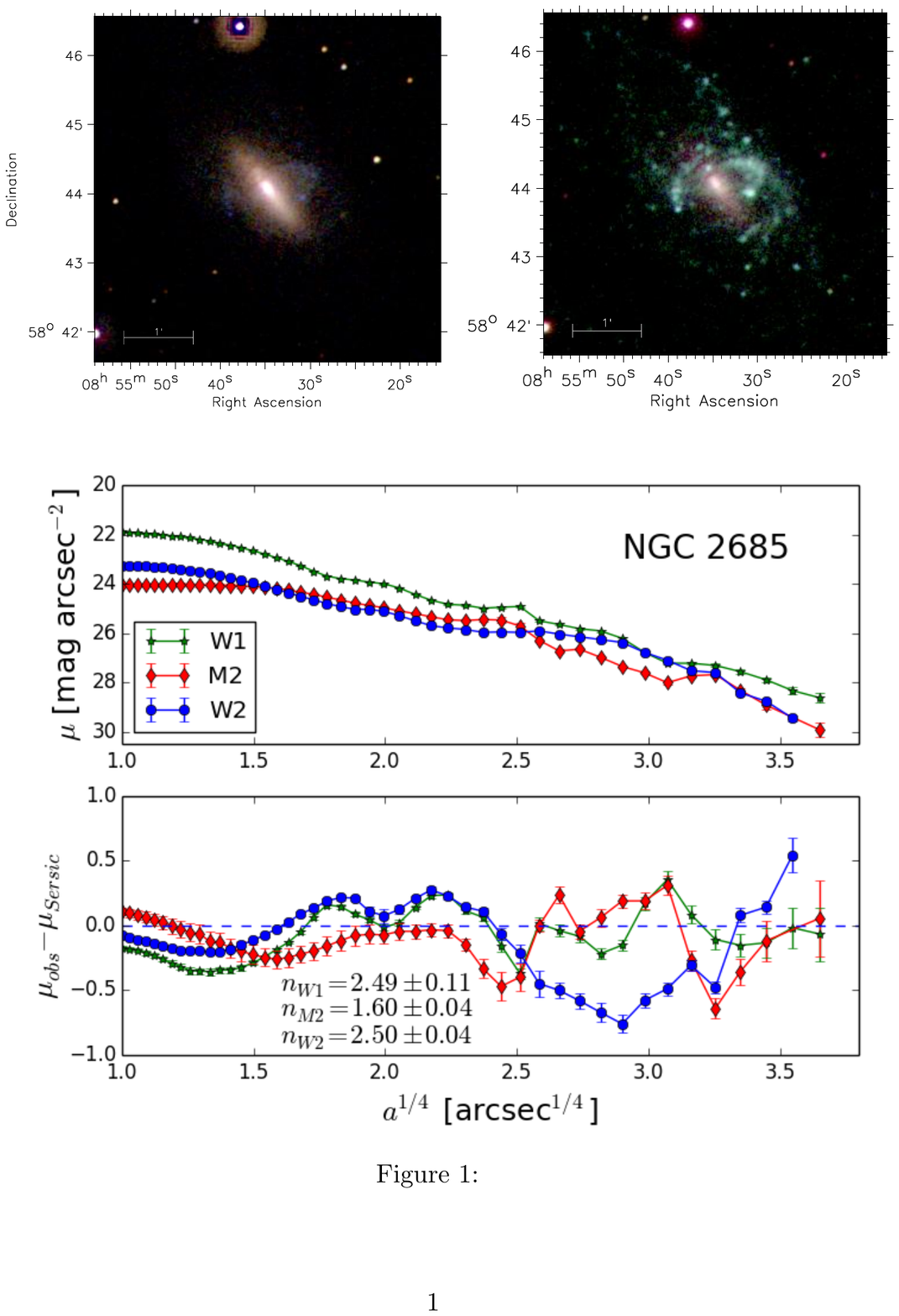}
	\caption{As in Figure~\ref{NGC1366-sb} for NGC~2685. The field of view is 
	5\arcmin$\times$5\arcmin.}
	\label{NGC2685-sb}
\end{figure*}

\begin{figure*} 
	\center
\includegraphics[width=15.5cm]{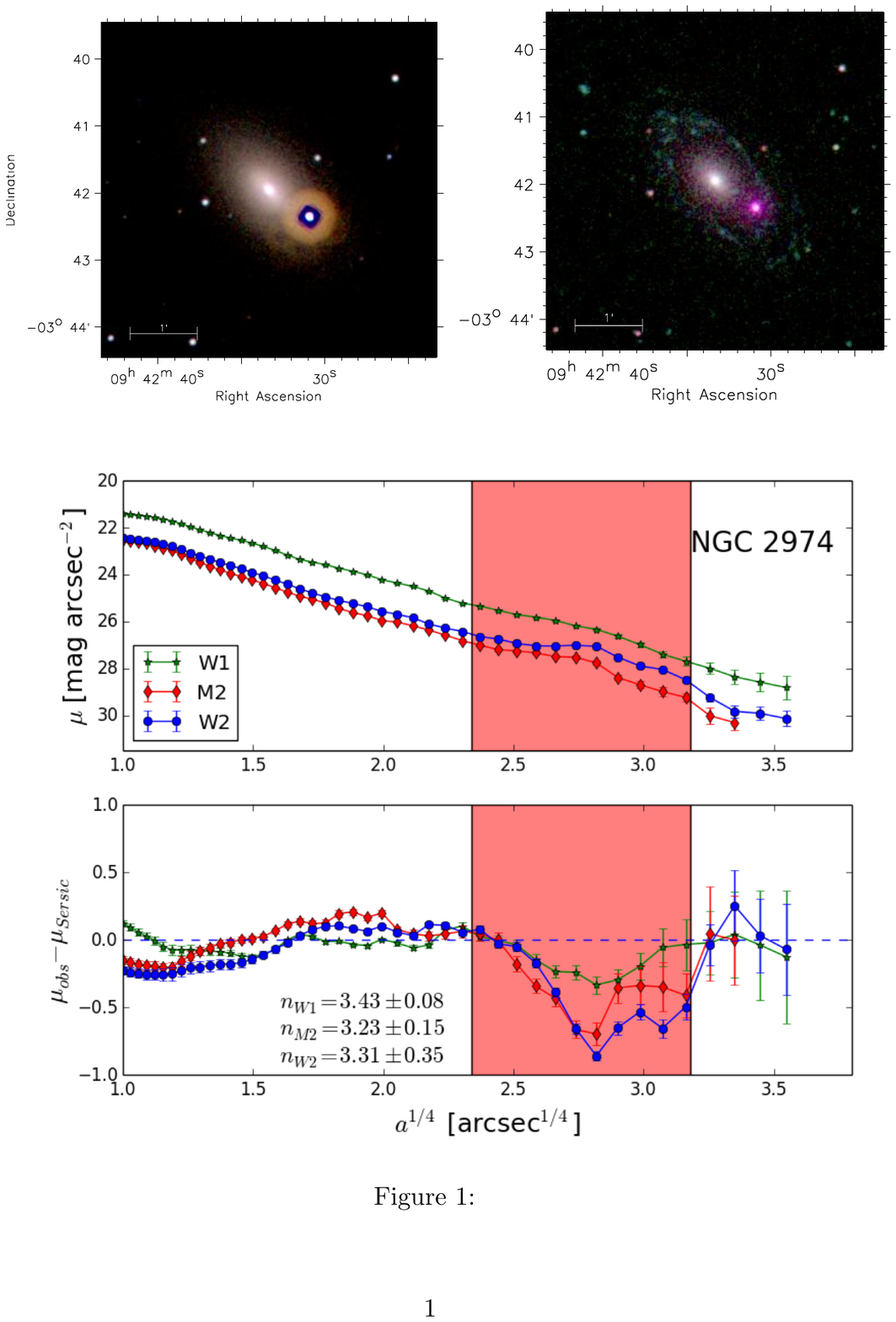}
	\caption{As in Figure~\ref{NGC1366-sb} for NGC~2974. The field of view is
	 5\arcmin$\times$5\arcmin. The shaded area indicates the masked region in the
	S\'ersic law fit (see text).}
	\label{NGC2974-sb}
\end{figure*}
\begin{figure*} 
	\center
\includegraphics[width=15.5cm]{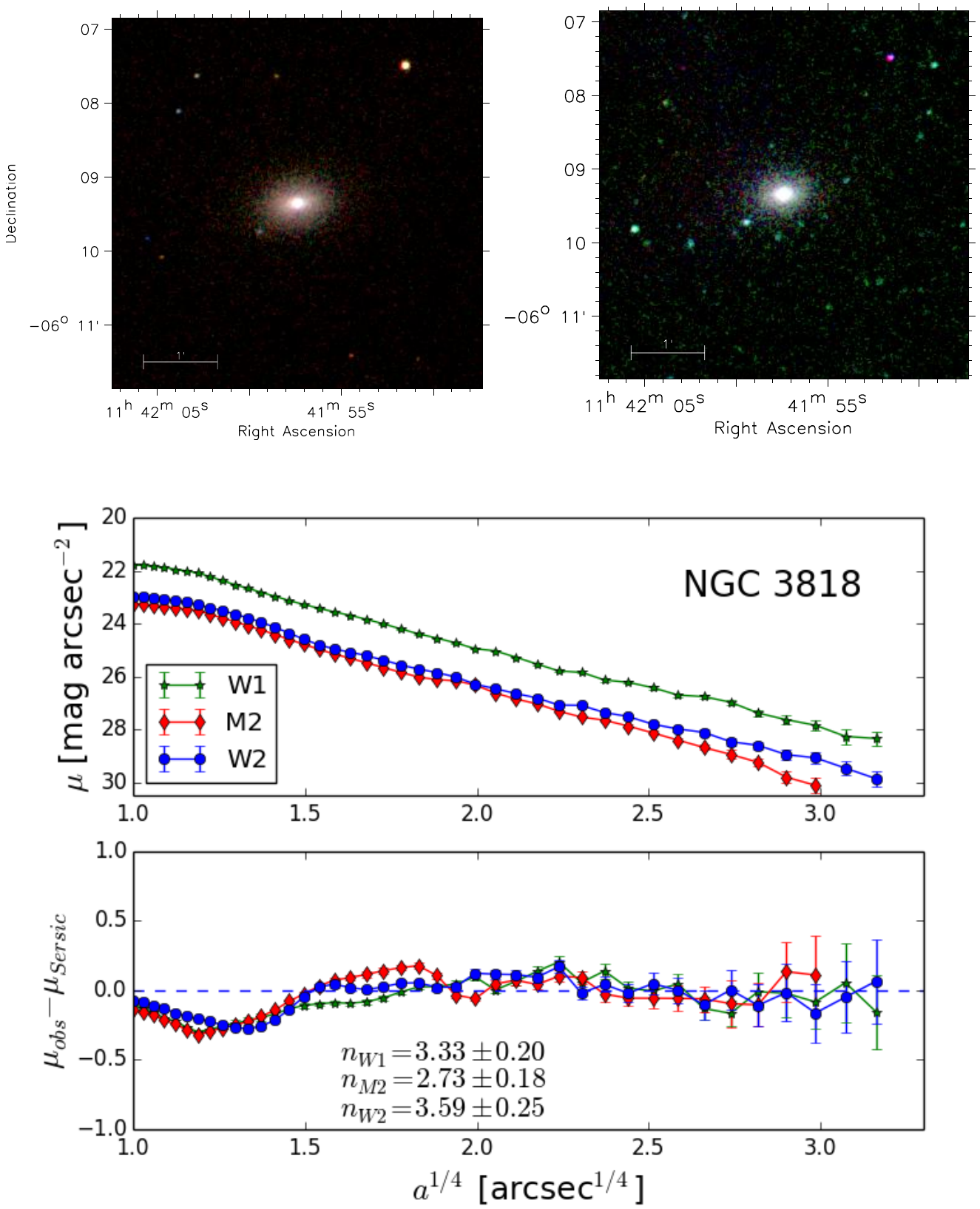}
	\caption{As in Figure~\ref{NGC1366-sb} for NGC~3818. The field of view is 
	5\arcmin$\times$5\arcmin.}
	\label{NGC3818-sb}
\end{figure*}

\begin{figure*} 
	\center
	\includegraphics[width=15.5cm]{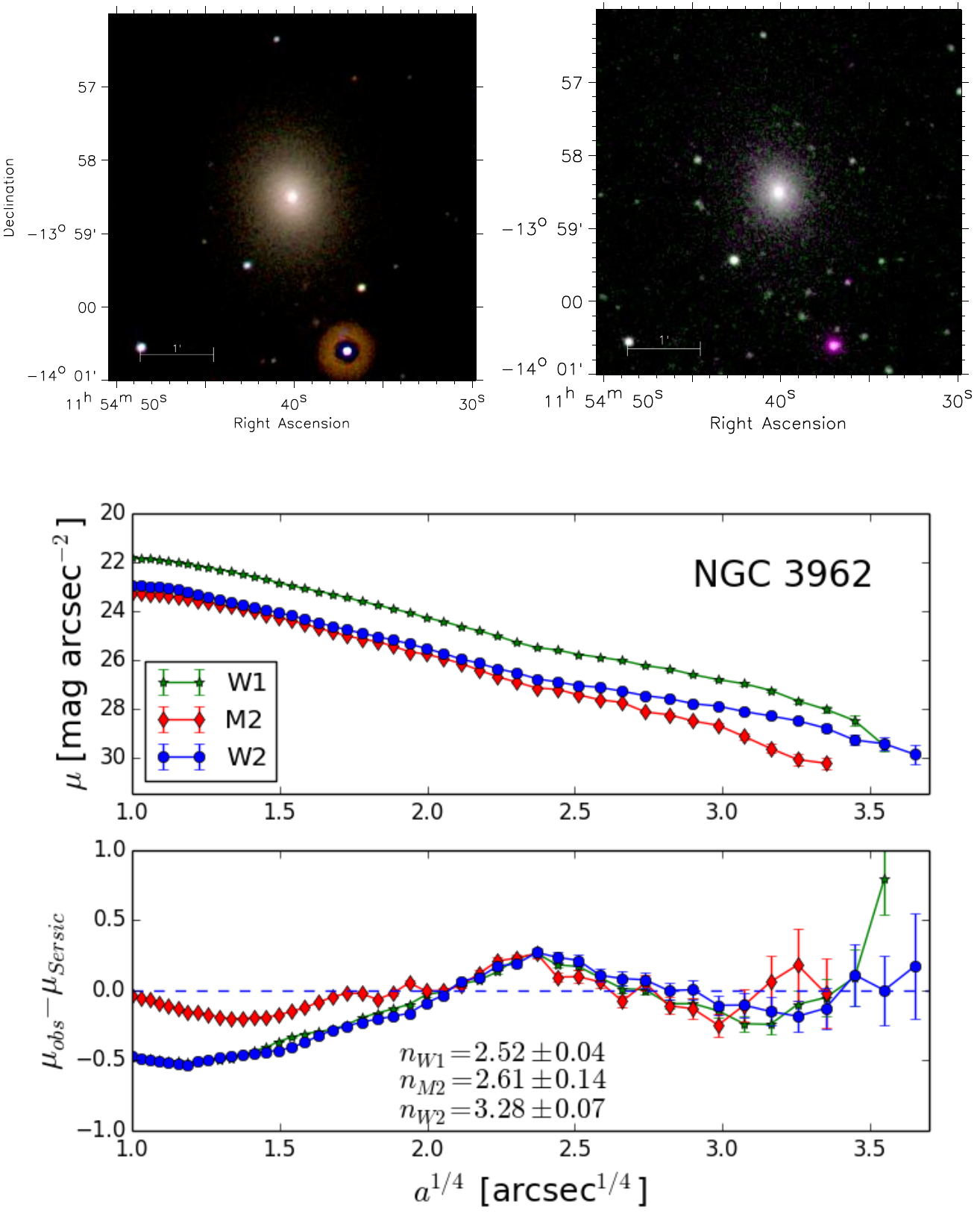}
	\caption{NGC~3962. As in Figure~\ref{NGC1366-sb}. 	The field of view is
	 5\arcmin$\times$5\arcmin.}
	\label{NGC3962-sb}
\end{figure*}

\begin{figure*} 
	\center
	\includegraphics[width=15.5cm]{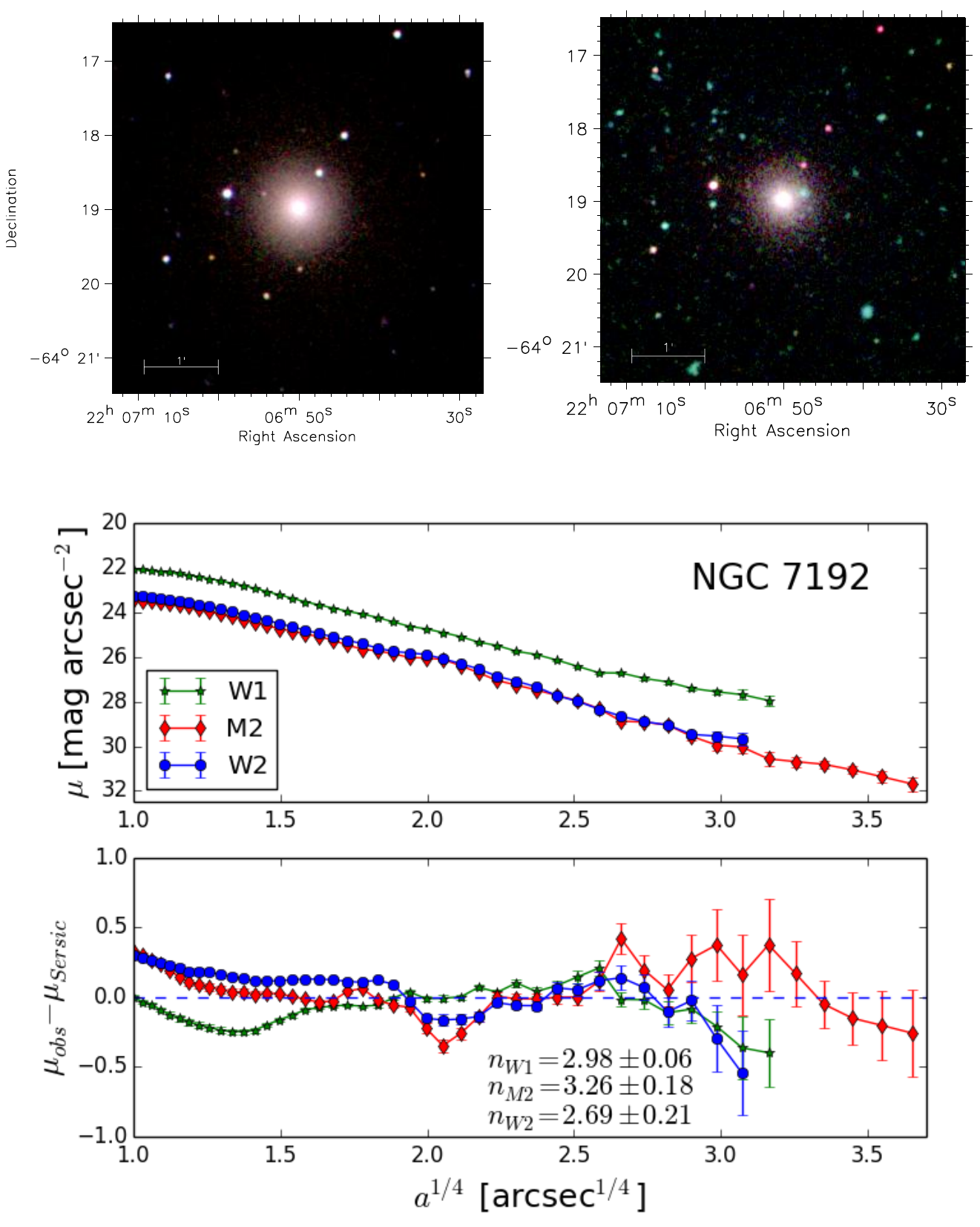}
	\caption{As in Figure~\ref{NGC1366-sb}  for NGC~7192. The field of view is 
	5\arcmin$\times$5\arcmin.}
	\label{NGC7192-sb}
\end{figure*}

\begin{figure*} 
	\center
	\includegraphics[width=15.5cm]{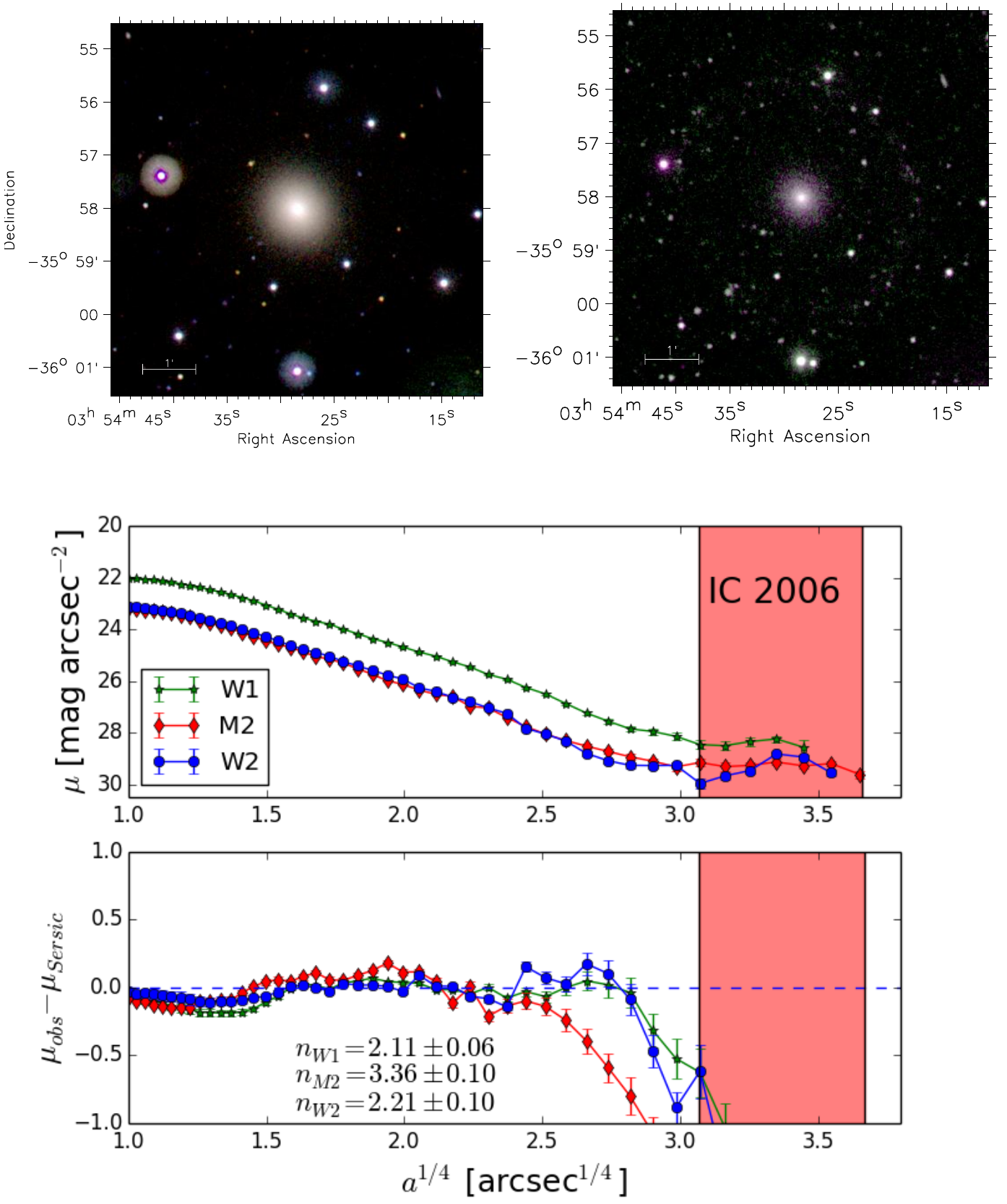}
	\caption{As in Figure~\ref{NGC1366-sb} for IC~2006.  The field of view is 
	7\arcmin$\times$7\arcmin. The shaded area indicates the masked region in the
	S\'ersic law fit (see text).}
	\label{IC2006-sb}
\end{figure*}

\medskip\noindent
\underbar{NGC~1415} ~~~  
Our photometry extends out to $\sim$16.5 kpc in radius, 
4$\times$r$_{eff}$ on average in our NUV bands. The optical and NUV 
 images (Figure~\ref{NGC1415-sb}, top panels) show
the complex structures present in the galaxy which justify its controversial 
classification: an S0 for the RC3 and a Spiral for the RSA. 
A complex system of dust is also visible in the optical
composite  image of  the CGS \citep{Ho11}.
More recently, \citet{Laurikainen11} classified the
galaxy as (RL)SABa (r'l,nr)0+ in their Near Infrared Atlas. 
\citet{Buta15} assign a morphological type T=-1
adding the notation "grand-design spiral".

In short, recent classifications tend to describe the galaxy as an 
S0 having an outer ring-lens (RL) system, without bar, with an inner ring-lens 
(r'l) and a nuclear ring (nr). The system of  rings
dominates our NUV images,  especially in the $M2$ and $W2$ filters,
and reflects on the shape of the luminosity profiles:  there is not a peak of luminosity
in the center since the luminosity of the nuclear ring prevails.  
The outer ring, well visible in the NUV bands,  shows a position 
angle different from the inner one as pointed out by \citet{Comeron14}.
We measure the position angle, PA, and, at the outer edge of the NUV features,
the semi-major, $a$ and semi-minor, $b$, projected axes. The nuclear ring 
(nr) has  PA=166$^{\circ}\pm2^{\circ}$, $a=10.8\arcsec$, $b=5.2\arcsec$;
the inner ring-lens (r'l) PA=139$^{\circ}\pm2^{\circ}$, $a=50\arcsec$, $b=14\arcsec$;
the outer ring (RL) PA=150$^{\circ}\pm2^{\circ}$, $a=152.8\arcsec$, $b=60.3\arcsec$.  

The ring/arm-like features perturb the entire NUV luminosity profiles.
Without masking any feature, the S\'ersic indexes  show a large variation  0.85 $< n <$ 2.46. 
The range of $n$ values highlights the presence of an underlying disk structure.

The B, V, R, I bands in the CGS cover an area radially more extended than our NUV profiles. 
The indices summarized in Figure~\ref{analysis-1} are the result of a fit  considering the CGS profile 
out to the NUV outermost observed radius. Moreover, the fit of the entire CGS profiles 
provides similar results, comparable within the errors. 
The S\'ersic index ranges from $2.83\pm0.04$ to $3.25\pm0.38$. Values of the 
S\'ersic index $n\geq3$ argue against  a disc morphology in spite of the fact 
that the galaxy shows a  ring revealing the presence of the disk.

The color profile, ($M2-V$) in Figure~\ref{color-M2-V}  shows
strong variations due to the presence of the ring/arm-like structures. 
The color is bluer in the inner regions notwithstanding the presence of dust clearly 
visible in our images, and redder in the outskirts.

\medskip\noindent
\underbar{NGC~1426}~~~
The galaxy is considered a {\it bona fide} elliptical by both the RC3 and the RSA catalogues.
The E classification was questioned by \citet{CPR88},   who
studied its geometrical structure  down to $\mu_B \approx$ 
28 mag~arcsec$^{-2}$. They suggested that NGC~1426 is  an S0 since 
the galaxy does not show any significant isophotal 
twisting $<PA>$ = 105$^\circ$ $\pm$1$^\circ$ out to
100\arcsec\ (PA=103.5$\pm$1.2 in \citet{Ho11})
as expected for disk galaxies. 
\citet{Huang13} model this galaxy with 3 S\'ersic components with  $n=2.17$, 0.69, 2.13. 
There  is no comment about a possible galaxy re-classification in their Table~1
on the basis  of   such low $n$ values. \citet{Hopkins09} provided
a S\'ersic fit of  the V-band optical profile which combines high resolution
 inner galaxy regions from {\it HST} and ground based measurements from different 
 sources. They propose a two component fit, "S\'ersic + cusp",   with a  value of the 
 S\'ersic index  5.26$^{+0.11}_{-0.35}$ (see their Table~1).

Our integrated NUV magnitudes (Table~\ref{tab5}) agree well with those in \citet{Hodges2014}.
The bottom panel of Figure~\ref{NGC1426-sb}  shows that
our  NUV luminosity profiles do not present peculiar
features. The range of  indexes, obtained with a single S\'ersic fit out to  $a\sim90$\arcsec,  is
$2.65 \leq n \leq 2.86$.

The CGS surface photometry extends out to $\sim$230\arcsec, although with large photometric
errors in the outskirts,  well beyond our NUV data.
The value of our S\'ersic fit of the CGS data, over the same range as our NUV analysis,
outlines values  lower than the classic $r^{1/4}$ law, just as in the  NUV bands (Figure~\ref{analysis-1}). 

The ($M2-V$) color profile of this galaxy (Figure~\ref{color-M2-V}) is rather flat.
The $a_4/a$ isophotal shape profile in the $M2$ band is very noisy in the outskirts, 
and shows some boxy isophotes in its intermediate part.   In the optical \citet{Li11} 
show $a_4/a$ (their $B_4$) consistent with 0 out to 100\arcsec.

\medskip\noindent
\underbar{NGC~1533}~~~
In the RSA catalog this  galaxy has an uncertain classification, barred S0/Spiral, while it is 
a barred S0 in the RC3 catalog.
The classification of \citet{Laurikainen11}, (RL) SB0$^{\circ}$, confirms the RC3 classification,
and  indicates the presence
of an outer ring lens (RL) associated to the barred S0 (Figure~\ref{NGC1533-sb}).
Rings are also discussed in \citet{Comeron14}.  A  S\'ersic index value  $n=1.5$ has 
been obtained by \citet{Laurikainen06}  with a 2D fit of  the bulge which
 they consider as a pseudo-bulge.
 {\it GALEX} evidenced a Far UV bright incomplete ring/arm-like feature
at  a radius of  $\sim 45''-55''$ \citep[see e.g.][and references therein]{Marino11c}
 with a  bright knotty structure.
This feature is also visible in  our  Swift  {\tt UVOT}  images.  
In particular, the top right panel of Figure~\ref{NGC1533-sb} shows 
that the UV bright structures are 
reminiscent of arms. Approximating the edge of the  UV bright North-West ``arc'' 
with an ellipse having   PA=120$^\circ\pm2$ and
semi-major and semi-minor axes $a=58\arcsec$ and 
$b=45\arcsec$, respectively,  we note that this
ellipse does not  include the  UV bright arm-like structure in the South-West.  
These arm-like structures  are  predominant 
features in the NUV luminosity profiles (Figure~\ref{NGC1533-sb}, mid panel).\\
\indent
The single S\'ersic fit, shown  in Figure~\ref{NGC1533-sb} (bottom panel),
 includes the NUV bright structures. We adopt these values of the fit 
and report them in Table~\ref{tab5} and in Figure~\ref{analysis-1}. 
 The residuals from the fit  emphasizes the regions 
covered by the ring/arm-like features, seen as excesses of light. 
The nucleus itself shows an excess of light with respect to the fit.\\
\indent
The bar is not as evident in the NUV (see also the $a_4/a$
profile in Figure~\ref{a4M2}) as it is in the optical bands.
In order to map the underlying galaxy structure we decided
to mask the CGS   $B$,  $V$, $R$, $I$  luminosity profiles in the region
from 20\arcsec\ to 90\arcsec, corresponding to the bar and the lens,
obtaining the  $n$ values shown in Figure~\ref{analysis-1}.  
We test the robustness of the range of the S\'ersic index variation
repeating the  fit of the NUV bands masking the region from 20\arcsec\ to 90\arcsec\ 
as in the optical wavelengths. The S\'ersic indices vary from 2.14$\pm$0.06 to  n=3.69$\pm$0.07,
in $W2$ and $W1$, respectively, i.e. the
values are similar to those in the unmasked NUV fits reported in 
Figure~\ref{NGC1533-sb}, so we used the unmasked Sérsic fit values for our analysis.  
Summarizing the values of the S\'ersic index, we select as the best representation 
of the underlying galaxy structure, range from $n\sim$2.54 ($M2$-band) to 4.89 ($V$-band).

The $(M2-V)$ color profile (Figure~\ref{color-M2-V}) becomes bluer with
increasing galactocentric distance beyond 16\arcsec. 
Knots and the arm-like structures discussed above appear
 bluer than the galaxy body. 

\medskip\noindent
\underbar{NGC~1543}~~~
The galaxy,  nearly face-on, is classified as a barred S0 
both in the RSA and the RC3. \citet{Laurikainen11}
classified the galaxy as  (R)SB(l,nl,nb)0$^+$ indicating the presence of a inner lens (l),
a nuclear lens (nl), a nuclear bar (nb) in addition to the bar remarked by the RC3 and
the RSA, and the outer ring (R). 
\citet{Erwin15} decomposed the galaxy luminosity profile with a S\'ersic law+exponential,
and reported the presence of a disk and of a composite bulge with $n=1.5$. 

The inner and outer bars as well as the lenses
 are visible in the {\tt UVOT} optical and NUV images. 
 The outer ring,   faint and diffuse in the optical images,  
 becomes brighter  in the NUV. Starting from  $U$ to $W2$  
a filamentary arm-like/ring structure emerges at the outer edge 
of this  ring (Figure~\ref{NGC1543-sb} top right panel). 

The NUV luminosity profiles  and  the values of   S\'ersic fit  are shown  in
the bottom panels of Figure~\ref{NGC1543-sb}. We exclude from the single S\'ersic
fit the  ring/arm-like structure in the galaxy outskirts. Notwithstanding, the fit is poor due
to the complex structure of the galaxy out to  its outskirts 
(see also the $a_4/a$ profile in Figure~\ref{a4M2}). The NUV S\'ersic indices  
are  lower than 3 and the  galaxy nucleus results brighter  than the S\'ersic 
model in the NUV filters. In the B, V, R, I bands the  S\'ersic fit shows that
the luminosity profiles are consistent with an r$^{1/4}$ law (Figure~\ref{analysis-1}).\\
The ($M2-V$) color profile shown in Figure~\ref{color-M2-V}
 reddens out to 57\arcsec\ where it starts to become bluer. Colors as blue as
 $M2-V\simeq$4 are found for the arm-like  structure  at r$>80\arcsec$.

\medskip\noindent
\underbar{NGC 2685}~~~The galaxy  is known as  the ``Helix" galaxy. RSA classifies
  this peculiar object, which shows polar 
rings and polar dust-lanes, as S0$_3$~pec.   
Its  polar rings/arcs/blobs, well known in the
optical bands, become quite spectacular in  NUV filters
(Figure~\ref{NGC2685-sb}, top right panel).\\
\indent
The NUV emission is very irregular in all  {\tt UVOT} filters. 
As a consequence, we performed a S\'ersic fit to the NUV 
luminosity profiles without masking any component
(see bottom panel of Figure~\ref{NGC2685-sb}).
Since optical luminosity profiles are not available in the literature,
we used  {\tt UVOT} $U$, $B$ and $V$ images to derive
 luminosity profiles and to estimate the S\'ersic indexes,
without masking any component as in the NUV. 
 Figure~\ref{analysis-1} shows the results: all the {\tt UVOT} bands have
very similar values,  $n\lesssim2.5$,  suggesting the presence of
an extended disk at all wavelengths. A deep optical view of this galaxy
is presented by \citet{Duc15} confirming the picture of an extended low surface
brightness disk and ring/arm-like structures (see the top left image in their Figure 17). 
Therefore, it is no surprising that the ($M2-V$) color of this galaxy 
becomes bluer, from $\sim$6 up to $\sim$2,  with  increasing 
 galactocentric distance (Figure~\ref{color-M2-V}).

\medskip\noindent
\underbar{NGC 2974}~~~ 
This galaxy is classified as  E4 in both the RSA and the RC3. However, 
\citet{Buta15} classified it as SA(r)0/a with type T=0.0
A bright star superposed on S-W side of the galaxy 
 hampers the study of this galaxy mainly at optical wavelengths.\\
\indent
The CGS optical image atlas shows  the presence of a faint extended ring/arm-like structure
in  NGC 2974 \citep{Ho11,Li11}. This structure  is very bright in the {\it GALEX} Far-UV  
\citep{Jeong09,Marino11a,Marino11b,Marino11c} and in
our  NUV images (top right panel of Figure~\ref{NGC2974-sb}).

The CGS optical and NIR luminosity profiles do not show signatures
of this structure which appears in our NUV luminosity profiles.
To describe the galaxy underlying
structure from NIR to NUV we considered two strategies. \\
We obtain the NUV S\'ersic indices both masking
the  luminosity profile from 30\arcsec to 100\arcsec\, where the
ring/arm-like features are prominent, and without masking,  as in the optical. 

In the first case the NUV S\'ersic  indices derived are
very similar to each other, ranging
from 3.23$\pm$0.15 in the $M2$ band to 3.43$\pm$0.10 in the $W1$ band
(these are reported in Figure~\ref{NGC2974-sb}, bottom  panel, and
in Figure~\ref{analysis-1}). In the second case, i.e. without any masking, we derive
 $n\lesssim2.5$ suggesting that the NUV emission marks a disk structure. 

The S\'ersic indices in NIR and optical bands from the CGS profiles
 range from 2.9  to 4.2 (see Figure~\ref{analysis-1}). 
\citet{Hopkins09}, fitting a cusp+S\'ersic laws, report a   value of 
the S\'ersic index of 4.06$^{+0.77}_{-0.48}$ in the V-band. 

The blue arm-like structure starts to emerge at $\approx$30\arcsec\
in the ($M2-V$) color profile of  Figure~\ref{color-M2-V}.

\underbar{NGC~3818}~~~ This galaxy is  classified E5 in the 
RSA and the RC3. A more detailed description
is given by \citet{Scorza98}  that consider NGC~3818  a bulge-dominated ETG,
having a disk fully embedded in a boxy bulge.

Our optical images  do not show remarkable features (top left panel
of Figures~\ref{NGC3818-sb}). The S\'ersic index  from NUV luminosity profiles is in the range
$2.73\pm0.18<n<3.59\pm0.25$. Since the galaxy does not belong to the CGS sample we use our
{\tt UVOT} data-set to study the optical wavelengths.

Our S\'ersic index estimate in the {\tt UVOT}-$V$ band is 3.09$\pm$0.17 (Figure~\ref{analysis-1})
in agreement with the value of \citet{Hopkins09}, $n=2.81^{+0.06}_{-0.06}$, in the same band.
 
We found that the galaxy is slightly disky, as shown in Figure~\ref{a4M2} and Table~\ref{tab5}.
Figure~\ref{color-M2-V} shows that this galaxy has a  red color profile, almost constant around 
an average value, ($M2-V$)$\sim$5.1.

\medskip\noindent
\underbar{NGC~3962} ~~~ The galaxy is classified E1 in the RSA and the RC3 and
E+3-4 by \citet{Buta15}. In the NUV bands we find that its  average  ellipticity 
is  in the range $\epsilon_{80}\sim 0.18 - 0.24$ (Table~\ref{tab5}).

The galaxy does not show an obvious disk structure, although 
the value of the $a_4/a$ isophotal shape profile tends to increase towards the
outskirts. Furthermore, the NUV 
luminosity profiles show two distinct trends with a break at $a^{1/4}$=2.45 arcsec$^{1/4}$.
This break, visible at the same radius also in the CGS luminosity profiles, 
is emphasized by our S\'ersic fit. S\'ersic index values
range from 2.52 to 3.28 in the NUV, and from 3.10 to 4.41 in the optical bands (see Figure~\ref{analysis-1}).
\citet{Huang13} find that the V-band profile is well fitted with 
three S\'ersic laws with indexes 3.25, 0.51 and 1.46.
The best fit of the sum of a cusp and a S\'ersic law in \citet{Hopkins09}  in the same  band gives $n=3.74$. 
The  2D single S\'ersic fit  of \citet{Salo2015}  is unsatisfactory ($n=6.1$) leaving
space for a better fit  that includes additional components.

Figure~\ref{color-M2-V} shows that this galaxy has a quite  red profile, constant around 
($M2-V$)$\sim$5.3 out to the galaxy outskirts.

\medskip\noindent
\underbar{NGC~7192}
The galaxy is classified S0 in the RSA and elliptical (.E+..*.) in the RC3. 
\citet{Huang13} suggest
that the galaxy is an S0 on the basis of a multiple S\'ersic fit. 
Their four component model recognizes
a bulge ($n=1.7$), two lenses ($n$=0.4 and $n$=0.5), and a disk ($n=0.9$).
The presence of two lenses in addition to an outer exponential disk 
has been previously  revealed by \citet{Laurikainen11}.   The \citet{Huang13} model and that
of \citet{Laurikainen11}   converge 
in indicating that this galaxy is an S0 rather than an E  \citep{Huang13}.\\
\indent
Our fits of the NUV luminosity profiles   (Figure~\ref{NGC7192-sb} 
bottom panel) provide values of $n$ slightly shallower ($n\approx3$) than the $r^{1/4}$ law.  
Similar  values are derived in the optical and NIR bands (Figure~\ref{analysis-1}).
The nucleus is  under-luminous respect to  the S\'ersic fit both in the $W2$ and $M2$ bands.\\
We point out that NUV images highlight the presence of 
 a blue knot at 17\farcs5  Nort-West from the galaxy center that 
does not correspond to any optical feature/star.  Possibly,
the knot is unrelated to the galaxy. The  ($M2-V$) color
profile  is almost constant around ($M2-V$)$\sim$5.3 (Figure~\ref{color-M2-V} ).

 \begin{figure*} 
 \center
 \includegraphics[width=12cm]{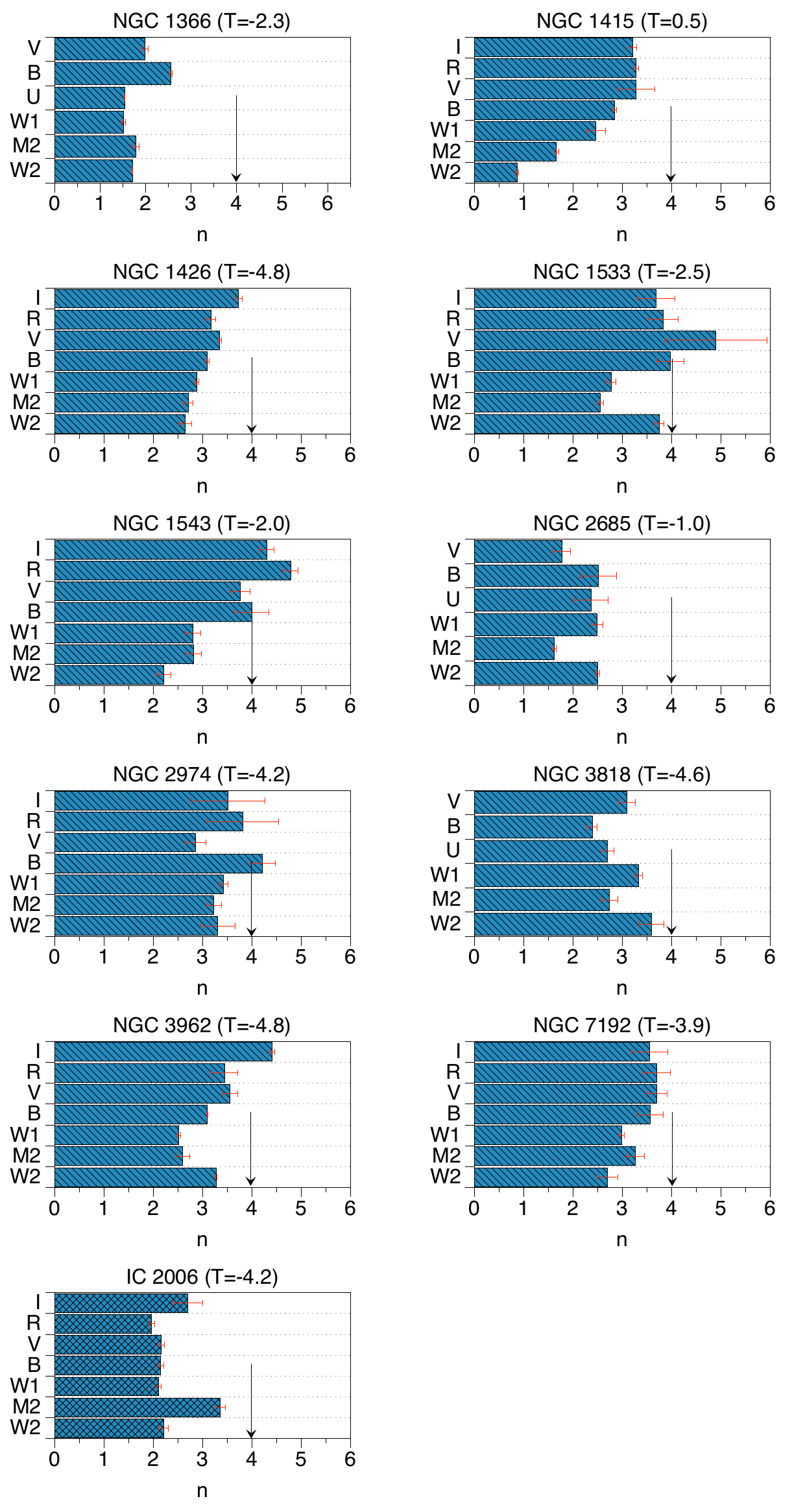}
 \caption{Synoptic view of the S\'ersic indices of the galaxies from NIR to  NUV  
 bands. NIR and optical S\'ersic indices  are obtained from the fit of the luminosity profiles 
  of the CGS \citep{Ho11,Li11}, except for NGC 1366, NGC 2685 and NGC 3818,  
  not included in that catalog. For these galaxies we used   {\tt UVOT} $U$, $B$ and $V$ bands. 
  The morphological type, T, is taken from {\tt Hyperleda} as reported in Table~\ref{tab1}.
The arrows indicate the r$^{1/4}$ law.}
 \label{analysis-1}
 \end{figure*}

\medskip\noindent
\underbar{IC~2006} ~~~
The galaxy is classified E1 both in the RSA and \citet{Buta15}, and  lenticular with 
an outer ring (RLA.-..)  in the RC3. The B-band  deep surface
photometry by  \citet{Schweizer} revealed indeed a faint ring. These authors,
however, suggest that the entire profile, including the ring which corresponds
to 1-2\% of the integrated B-band luminosity, is well approximated 
by a $r^{1/4}$ law, i.e. the galaxy is a "bona fide" elliptical.
\citet{Schweizer} found that the  ring overlaps  with an HI counter-rotating ring 
at a radius of about 150\arcsec,  inclined about 37.5\degr$\pm$2\degr 
with respect to the plane of the sky. \\
\indent
In our images the outer ring  starts to be recognizable in the $U$-band  
and becomes more prominent towards $W2$ (Figure~\ref{IC2006-sb},
top panel).  We measure the semi-major and minor axes of the  bright UV emission,
a=151\arcsec, and b=129\arcsec,  and the PA=$40^\circ \pm2^\circ$.\\ 
\indent
The presence of the ring reflects on the NUV luminosity profiles as shown in the
bottom panel of Figure~\ref{IC2006-sb}. In NIR-optical CGS luminosity profiles
this feature is not detected. In this context, we decide to exclude the ring,  and to fit the
S\'ersic law of our NUV profiles out to $90$\arcsec. This produces $n\leq3.3$
in the NUV bands (bottom panel of Figure~\ref{IC2006-sb}).  Our fit with a single S\'ersic 
law  out to 90\arcsec in the  $B$, $V$, $R$ and $I$ profiles in the CGS gives $n<$2.7 for all of them
(Figure~\ref{analysis-1}) and the same values i.e. $n<$2.7 fitting the entire profiles.\\
\indent
Summarizing, S\'ersic indices  in the range $n\sim2$ to $n\sim3.3$ 
are  found from NIR-optical to NUV, as shown
in Figure~\ref{analysis-1}.   A somewhat larger value, $n=3.5$, has been recently obtained 
by \citet{Salo2015} analyzing MIR-images of IC~2006. Our analysis suggests
than the ring/arm-like structure should lie on  an disk, although the global 
behavior of the $a_4/a$ does not show a clear evidence of such an underlying feature. \\
\indent
The ($M2-V$) color profile of this galaxy
 is constant out to 40-45\arcsec ( Figure~\ref{color-M2-V}) and  becomes  progressively bluer as the
galactocentric distance increases, well before 90\arcsec\ where the outer bright NUV ring emerges. 

 \begin{figure*} 
 \center
  \includegraphics[width=17.5cm]{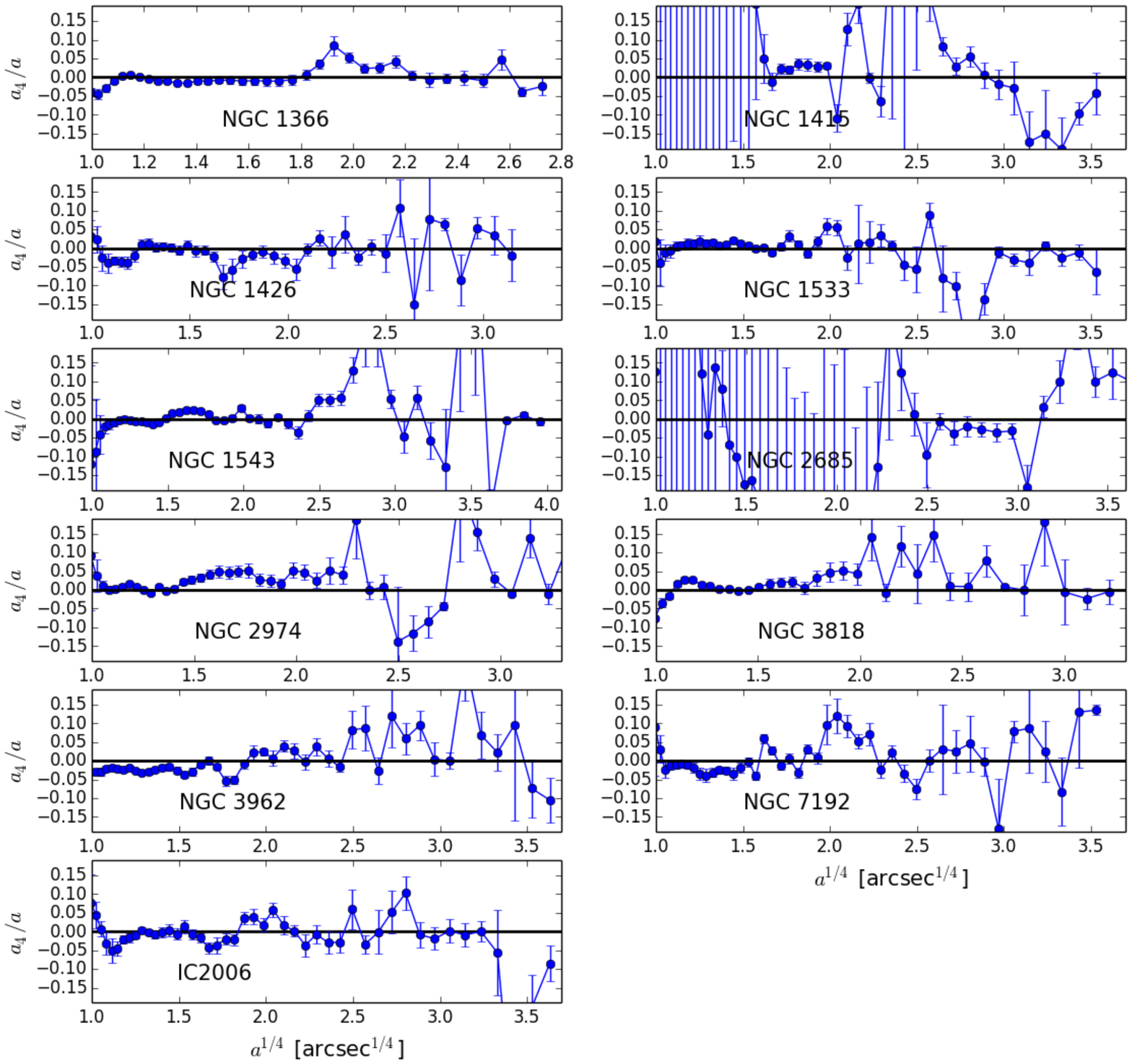}
 \caption{The isophotal  shape, $a_4/a$, as a function of the 
 semi-major axis in the $M2$ band (see also its average values
 in Table~\ref{tab5}).}
 \label{a4M2}
 \end{figure*}

 \begin{figure*} 
 \center
  \includegraphics[width=17.5cm]{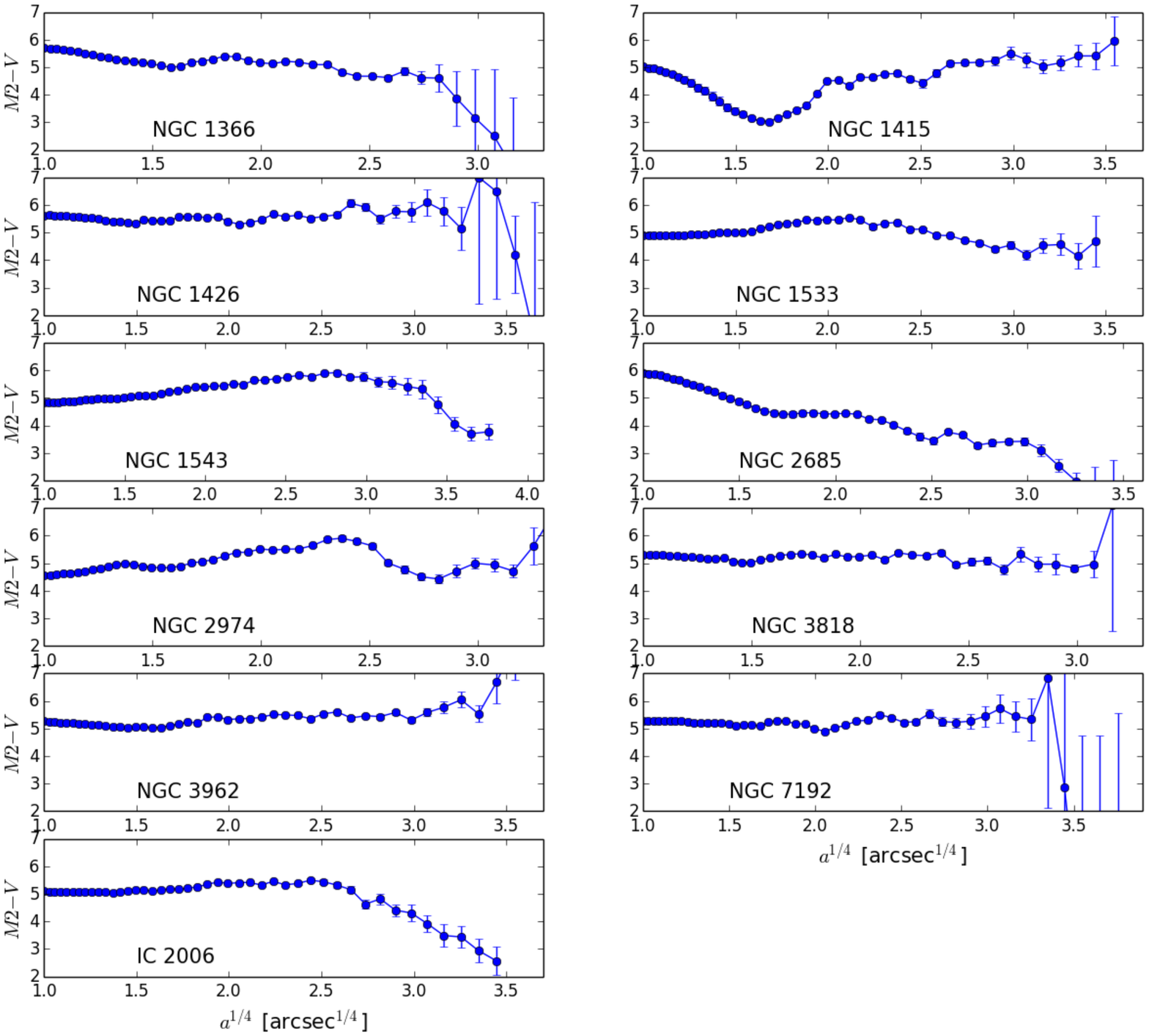}
 \caption{($M2$-V) color profiles of our ETGs sample. Profiles are corrected for the galactic extinction following
 \citet{Roming09}.}
 \label{color-M2-V}
 \end{figure*}

\section{Discussion}
\label{Discussion}
In the introduction we asked two questions: what can our NUV investigation tell us about the history 
of these ETGs, and whether  a common evolutionary framework exists for galaxies showing 
 so different NUV properties. With eleven galaxies only we cannot present a
coherent or exhaustive picture, in particular since we have a variety of behaviors and
the properties that apply sometimes to only one of the galaxies and not the others. A more general
discussion will have to await for more examples. Nonetheless, we can already see a few common features:
the indication that some recent star formation can be detected in the NUV frames, and confirmed 
by the S\'ersic index analysis, suggesting that {\it wet} accretion could be more common than
previously thought even in ETGs.
In the following section we connect our NUV results to the multi-wavelength 
information available in the literature for these galaxies to explore their evolutionary framework.

\subsection{The NUV galaxy structure}

Following the multi-wavelength S\'ersic index analysis developed in the
literature from NIR to $u$ bands
\citep{LaBarbera2010,Vulcani14,Kennedy16} we use 
our small sample to extend the investigation to the NUV regime. 

Figure~\ref{analysis-1} summarizes the values of the S\'ersic indices we derived,
from NIR to NUV. As described in the previous section, for each galaxy the 
fit  has been performed with the aim of obtaining  the description
of the underlying galaxy structure in the different bands.
The figure shows that most of the NUV S\'ersic indexes are, 
within errors, in the range 2-3 or  lower  whereas values corresponding to optical and 
 NIR bands tend to be larger, $n\approx3-4$. 
Based on the evidence provided by NGC~1366, where a edge-on disk 
is clearly visible in the NUV images, and all the bands but one have $n<2$  
we can interpret the low $n$ values as evidence of
 a disk structure in the NUV, as already done  at longer wavelengths.  
 Therefore, we can attribute a NUV disk to IC~2006 and NGC~2685.

When comparing the S\'ersic index values within each individual galaxy, we notice 
that variations are less extreme in Es (NGC~2974 and NGC~3818,  T$<-4$).
The widest range, consistent with the trend
expected for spiral galaxies, is found in NGC~1415, with the "latest" morphology  (T=0.5).  
We therefore, can conclude that, in spite of the small size of the sample,
we can reproduce the basic results found in the larger samples in 
optical and NIR bands: 
Es, differently from later types,  do not show a significant variation in the S\'ersic indices 
with wavelengths \citep{LaBarbera2010,Vulcani14,Kennedy16}. However, 
with our extension in the NUV realm, S\'ersic indices suggest 
a disk component in ETGs, likely connected to a recent accretion history and a recent
star formation activity. \\
\indent
{\it Do the color profiles indicate the presence of a disk with color bluer than the bulge?}
  Figure~\ref{color-M2-V} shows  that there is  a large variation in the
($M2-V$) color profiles of our sample.
Four galaxies namely  NGC~1426,
NGC~3818, NGC~3962,  and NGC~7192,  have nearly constant colors. 
For them there is no evidence of a disk that would make the profile bluer with 
increasing radius. 
Mazzei et al. (2017, in preparation) will show, in simulating our ETGs,
 that the color of the younger galaxy populations 
becomes as red as (M2-V)$\approx$ 5.5 in few $10^8$ yr. 
This behavior, i.e. a global reddening of the galaxy with age, has been found
in other bands by \citet{Kennedy16}. They found that both the bulge and the 
disk become redder with redder total color.
We note in addition that the isophotal shape  (Figure~\ref{a4M2})  suggests 
a disky appearance at the outskirts of both NGC~3818 and NGC~3962. \\
\indent
 We notice that the color of NGC~1366, NGC~1533, NGC~1543, NGC~2974 and
IC~2006 becomes blue at large radii.  
The color profile of NGC~1415 is bluer in the intermediate region than it is 
in the outskirts, likely due to the presence of dust in the outskirts. 
Only in NGC~2685 the color  profile becomes nearly monotonically bluer with radius.\\
\indent
Summarizing, although the S\'ersic analysis of  NUV luminosity profiles with $n\approx2-3$
suggests the presence of an underlying NUV disk independently of the morphological class 
and the presence of ring/arm-like structures, the behavior of ($M2$-$V$) color shows
a variety of behaviors. Indeed, the color provides less compelling evidence since it
can vary rapidly.

\subsection{A "wet" evolutionary scenario}

The presence of a disk highlighted by our NUV analysis
requires a dissipative origin i.e. a recent "wet' 
galaxy evolution scenario. {\it Is this scenario supported by other observations? 
What are the drivers of this evolution?} 
To address the above questions we choose to investigate SF and
kinematical properties dividing ETGs with an unperturbed morphology 
(NGC 1366, NGC 1426, NGC 3818, NGC 3962 and NGC 7192) from those showing
ring/arm-like structures (NGC 1415, 
NGC 1533, NGC 1543, NGC 2685, NGC 2974 and IC 2006). 

In our sample there are several signatures of star formation events even in galaxies 
with old or relatively old luminosity weighted nuclear ages (e.g. r$_e/8$)  \citep{Annibali07}.  
\citet{Amblard14} reported a difference 
 of  a factor $\approx$55 between the star formation rate (SFR) of NGC~2685, a polar ring galaxy
with the highest star formation rate in our sample (log SFR=-0.36$\pm$0.41 $M_\odot\ yr^{-1}$), 
and that of the nearly unperturbed galaxy NGC~1426  (log SFR=-2.08$\pm$0.15 $M_\odot\ yr^{-1}$).
{\it Spitzer}-IRS observations \citep{Rampazzo13} revealed that NGC~1533, NGC~2685,
NGC~2974 and NGC~3962 show nuclear  PAHs, tracers of some star formation 
events in the past 1-2.5 Gyr  \citep[see e.g.][and references therein]{Vega10}.\\  
\indent
Concerning evolutionary drivers, there is a  {\it crescendo} of signatures, 
going from unperturbed morphologies to those characterized by 
NUV ring/arm-like structures, supporting  a "wet" accretion scenario 
in our ETGs. First of all the fundamental ingredient for a "wet" evolution, the neutral gas, is still 
abundant. Many of our ETGs own significant HI gas reservoirs (see column 8 in Table~\ref{tab1})
that can feed star formation episodes. In addition, kinematical peculiarities
point towards interaction/accretion events.
\medskip

Among ETGs with an unperturbed morfology, NGC 1426 and NGC 3818, 
do not have indication of accretion episodes. \citet{Annibali10} report that emission 
lines in the optical are weak in  NGC 3818 and absent NGC 1426, supporting, 
together with the class 0 of their MIR spectra \citep{Rampazzo13}, the passively
evolving  nature of the stellar populations in their nuclear region.
Moreover, \citet{Annibali07} reported an old luminosity weighted age for NGC 1426
(9.0$\pm$2.5 Gyr) and for NGC 3818 (8.8$\pm$1.2 Gyr) from line-strength
indices analysis. \\
\indent
In the remaining ETGs with an unperturbed morphology,   namely
NGC~1366, NGC~3962 and NGC~7192, signatures of accretion are found.  
In these galaxies a past star formation episodes may have {\it rejuvenated} the 
nucleus \citep{Annibali07,Rampazzo13}. \\
\indent
The nucleus of NGC~1366 has a relatively  young 
luminosity weighted age of 5.9$\pm$1.0 Gyr within r$_e$/8 
found by \citet{Annibali10}. 
 \citet{morelli}  found that NGC 1366 hosts  a nuclear kinematically decoupled component
younger than the host bulge. To explain the properties of the counter-rotating component,
 \citet{morelli} suggested  that enriched
material has been recently acquired via interaction or minor merging.
NGC~7192 also has  a relatively young nucleus (5.7$\pm2.0$\,Gyr) 
\citep{Annibali10}.  \citet{Tal09} detected a shell system around this
galaxy, witnessing its past accretion history.
 \citet{CD94} showed that stars in the innermost 8\arcsec\,region counter-rotate 
with respect to stars at larger radii. 

NGC~3962 is a particularly intriguing case.
The nuclear luminosity weighted age of its stellar populations  is 10.0$\pm$1.2 Gyr 
\citep{Annibali07}.
However, the nucleus hosts a radio source \citep{Bi85,Brown11},
and recent star formation, since there are PAHs and 
emission lines (MIR  spectral class 2,  \citet{Rampazzo13}).  
In addition the arm-like  structure detected by
\citet{Buson93} in H$\alpha$+[NII] supports the presence of a recent/on-going
star formation episode \citep[][report a log SFR=-1.32$\pm$0.20 $M_\odot\ yr^{-1}$]{Amblard14}. 
The NUV and CGS optical luminosity profiles of NGC 3962 show a sudden change of slope 
at $\approx$35\arcsec.  This behavior is remarkably 
different from the luminosity profiles of  the other objects 
 in our sample  (see Section~\ref{Individual Notes}).
Two distinct subsystems are also detected by  \citet{Z96}: an inner gaseous 
disk hosting an old stellar bulge, and an arm-like structure. 
The inner disk shows regular kinematics with a major
axis near PA= 70\degr\ and an inclination of about 45\degr\ and a radius of
 about 5\arcsec, at the limit of our, better than  {\it GALEX}, resolution. 
The brighter arm-like feature extends about 20\arcsec corresponding to 3\,kpc
(Table~\ref{tab1}). The H$\alpha$+[NII] emission \citep{Z96}, however, does not extends
out to the region where the change of  the slope of the NUV luminosity profile occurs.
Neither our  {\tt UVOT}-NUV study nor {\it GALEX} observations \citep{Marino11c}
show  NUV features associated to the H$\alpha$ emission.

 \begin{figure} 
 \center
 \includegraphics[width=8.5cm]{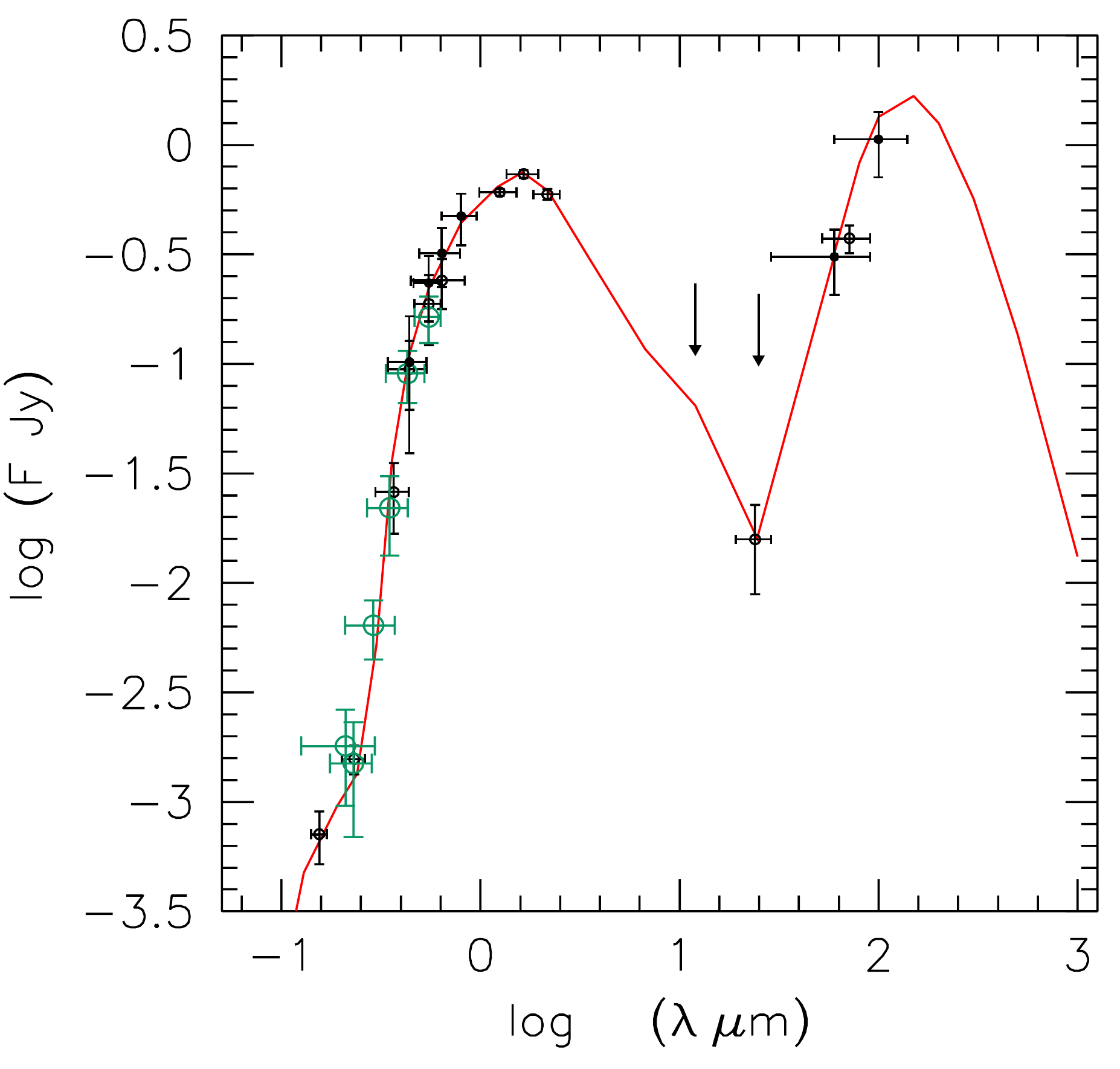}
 \caption{SED of NGC 1533 with {\tt UVOT}-NUV data (green open circles).  
 The red solid line is the predicted SED, and black open circles and upper limits are observed data 
 as in \citet{Mazzei14b}.}
 \label{sed-NGC1533}
 \end{figure}

Signatures of  "wet" accretion/merging episodes are more evident
in ETGs with ring/arm-like structures.  \\
\indent
There are no direct signatures of accretion in NGC 1415. 
The PA of the outer ring (RL), well visible in our NUV images (see Figure~\ref{NGC1415-sb} 
top panels), differs from that  of the inner ring (r'l)  as reported by \citet[][]{Comeron14}. 
These authors found that the decoupling between
the position angles  of the inner and outer rings is quite normal among ringed galaxies.  
However, NGC 1415 is located in the Fornax-Eridanus cloud (Table~\ref{tab1} column 9).
Since Eridanus is a loose, irregular, low velocity dispersion group composed of 31 members, 
many of which spirals \citep{Brough2006}, one cannot discard the hypothesis that  the 
decoupling observed in NGC 1415 may be an effect of  the interactions within the group.\\
\indent
NGC~2685, a polar-ring galaxy with  gaseous and 
 stellar components dynamically decoupled, is the most striking example
for evidence for wet accretion.
The outer stars rotate on a polar orbit about a central 
lenticular galaxy  \citep{Whitmore90}. \citet{Ulrich75} 
detected an ionized gas component  rotating fast perpendicularly to the 
galaxy major axis \citep{Schechter78}.
The kinematic separation of the two systems is also visible in HI 
\citep[][and references therein]{Jozsa09}. \\
\indent
NGC~2974, imaged in H$\alpha$,  presents 
some peripheral fainter filaments \citep{Busonetal93}  and an ionized disk \citep{U84,Gou94} 
misaligned with respect to the stellar isophotes by $\approx$ 20\degr  \citep{Pi97}.
NGC~2974 also has  a HI disk  \citep{Kim88}  with the same rotation axis as the stars. 
\citet{Kilborn09} revealed HI emission  in both NGC~1533 and NGC~1543
(their Figure 4). A detailed map of the  very extended HI distribution around  NGC~1533 
has been  provided by \citet{Werk10}.  IC~2006 has an HI ring corresponding to the NUV 
emission as reported above \citep[see][]{Schweizer}. The HI is
counter-rotating with respect to the stars in the central
body of the galaxy. The presence of  two peaks of the stellar velocity dispersion
map, derived by  \citet{Scott14}  with the {\tt WiFes} spectrograph, can be  interpreted 
as a signature of a  second counter-rotating component in the inner part 
of the stellar velocity field. \\
\indent
Our team   already explored the formation mechanism and the evolution of ETGs in two nearby 
groups \citep{Mazzei14a} and of some ETGs with ring/arm-like structures 
simulating, in particular,  NGC~1533. We  use smooth particle hydrodynamic simulations 
with  chemo-photometric implementation starting from
halos of gas and dark matter \citep{Mazzei14b} to derive the dynamical and morphological 
evolution as well as the spectral energy distribution (SED) 
extending over four orders of magnitude in wavelength at each time.
In Figure~\ref{sed-NGC1533} the
{\tt UVOT}  integrated magnitudes of  NGC~1533, here derived,  are over-plotted 
(green open circles) to the predicted galaxy SED (red solid line). The new data agree well 
with  prediction by  \citet{Mazzei14b} and are consistent with their results concerning 
the evolutionary scenario of this galaxy, that is 13.7 Gyr old according to the
simulation, which suggests that NGC~1533 is the product of a wet merger following a  2:1 head-on collision. 
The UV ring, a transient feature in this galaxy evolution, appears 
when the galaxy is $\approx$8 Gyr old i.e. in the latter stages of the merger episode.\\
\indent
We conclude that multi-wavelength, kinematical and spectroscopic investigations accumulate
in indicating that interaction/accretion/merging
episodes have characterized the recent history of many of our galaxies. Many of such
 episodes left signatures of recent star formation, either in their nuclei and/or in the 
outskirts.  The star formation episodes  give evidence that accretions/mergers have been
"wet", i.e. that dissipation  is the mechanism at the origin of the 
underlying disk structure revealed by  our analysis of NUV luminosity profiles. 

\section{Summary and conclusion}
\label{Summary}

We present NUV  surface photometry of eleven ETGs observed
with {\it Swift-{\tt UVOT}}. These galaxies  had their X-ray properties 
presented in Paper~I.

We derive  their  integrated magnitudes and luminosity profiles 
in the $W2$, $M2$ and $W1$ NUV bands and as well  in  optical bands.

We find a variety of behaviors, ranging from the "expected" 
featureless NUV images in about half of the galaxies examined
(NGC~1366, NGC~1426, NGC~3818, NGC~3962 and NGC~7192)  to 
ring-arm/like structures, to evidence of a polar ring (NGC 1415, NGC 1533,
NGC 1543, NGC 2974, IC 2006, NGC 2685).

In particular: 

\begin{itemize}

\item 
We perform a homogeneous S\'ersic analysis of both our NUV and optical data
from {\it Swift} and  B, V, R, I luminosity profiles from the CGS  finding the following results.
For most of our ETGs  without and with ring/arm-like structures, regardless 
the optical classification,  S\'ersic indices in NUV are
on the average in the range $n\approx2-3$  suggesting the presence of an 
underlying disk. Excluding disk galaxies (e.g. NGC 1366), 
S\'ersic indices from optical profiles are in general larger than
those in NUV, being on the average in the range $n\approx3-4$.

\item
The ($M2$-V) color profiles evidence that the ring/arm-like structures 
are bluer than the galaxy body suggesting that they host on-going or recent star formation. 
Featureless ETGs have red colors, however,
in NGC 3962 and NGC 3818 the $a_4/a$ isophotal shape is disky
in the outskirts. We remind that the color of young stellar populations
turns to red ($M2$-V)$\approx$5.5 in a few $10^8$ years. 

\end{itemize}

The values of the S\'ersic indices in the NUV reveal the role of a dissipative mechanism
in these ETGs, which is clearly supported by the presence of residual HI gas. Also 
NUV unperturbed systems present scars
of accretion/merging episodes that "rejuvenated'' the galaxy nucleus, as indicated
by both their luminosity weighted nuclear ages and PAH observations suggesting star formation 
episodes in the last few Gyrs. Decoupled gas-stars kinematics further support this view.\\
\indent 
We plan to use these far UV data-set, together with the X-ray analysis already 
presented in Paper~I, to constrain a grid of SPH simulations with chemo-photometric 
implementation providing a self-consistent view of the  evolution of our ETGs.
 In Figure~\ref{sed-NGC1533} we show how the basic information collected 
here about their NUV properties  will help in defining the SED of  NGC 1533 already 
studied in \citet{Mazzei14b}.  \\
\indent
We further emphasized that {\it Swift}-{\tt UVOT} data represent 
an important NUV resource for the study of ETGs in particular for those galaxies which 
do not have deep {\it GALEX}  images.

\section*{Acknowledgments}

We thank the anonymous referee for the careful reading and helpful suggestions 
that improved the paper.
We would like to thank Louis Ho, Zhao-Yu Li and the CGS team for having 
kindly provided us with B, V, R, I luminosity profiles derived 
form the Carnegie-Irvine Galaxy Survey. Paola Mazzei and 
Roberto Rampazzo acknowledge support from INAF through grant 
PRIN-2014-14 `Star formation and evolution in galactic nuclei'.
We acknowledge the usage of the {\tt HyperLeda} database 
(http://leda.univ-lyon1.fr).

\label{lastpage}


\begin{thebibliography}{99}

\bibitem[\protect\citeauthoryear{Amblard et al.}{2014}]{Amblard14} Amblard, A., Riguccini, L., 
Temi, P., Im, S. Fanelli, M., Serra, P. 2014, ApJ, 783, 135 

\bibitem[\protect\citeauthoryear{Annibali et al.}{2007}]{Annibali07} Annibali, F., Bressan, A., 
Rampazzo, R., Zeilinger, W. W., Danese, L. 2007, A\&A, 463, 455

\bibitem[\protect\citeauthoryear{Annibali et al.}{2010}]{Annibali10} Annibali, F., Bressan, A., 
Rampazzo, R., Zeilinger, W.~W., Vega, O., Panuzzo, P. 2010, A\&A, 519, A40

\bibitem[\protect\citeauthoryear{Bender et al.}{1989}]{Bender89} Bender, R., Surma, P., D\"obereiner, S. M\"ollenhoff C., 
Madjewsky, R. 1989, A\&A, 217, 35

\bibitem[\protect\citeauthoryear{Birkinshaw \& Davies}{1985}]{Bi85} Birkinshaw, M., Davies, R.L.  1985, ApJ 291, 32


\bibitem[\protect\citeauthoryear{Breeveld et al.}{2010}]{Breeveld10} 
Breeveld, A.A., Curran, P.A., Hoversten, E.A, Koch, S., Landsman, W. et al. 2010, MNRAS, 406, 1687

\bibitem[\protect\citeauthoryear{Breeveld et al.}{2011}]{Breeveld11} Breeveld, A.A., Landsman, 
W. 
Holland, S.T. et al. 2011, in GAMMA RAY BURSTS 2010. AIP Conference Proceedings, 1358, 
373

\bibitem[\protect\citeauthoryear{Bressan et al.}{2006}]{Bressan06} Bressan, A., Panuzzo, P., Buson, L., Clemens, M., Granato, G. L., Rampazzo, R., Silva, L., Valdes, J. R., Vega, O., \& Danese L. 2006, ApJ, 639, L55

\bibitem[\protect\citeauthoryear{Brown et al.}{2011}]{Brown11} Brown, M.J.I., Januzzi, B.T., Floyd,  D.J.E., Mould, J.R. 2011, ApJL, 731, L41

\bibitem[\protect\citeauthoryear{Brough et al.}{2006}]{Brough2006} 	
Brough, S., Forbes, D. A., Kilborn, V. A., Couch, W., Colless, M. 2006, MNRAS, 368, 1351


\bibitem[\protect\citeauthoryear{Burrows et al.}{2005}]{swift3}
Burrows, D. N., Hill, J. E., Nousek, J. A., et al. 2005, Space Sci. Rev., 120, 165

\bibitem[\protect\citeauthoryear{Buson}{1993}]{Buson93} Buson, L. 1993, Mem. SAIt, 64, 629

\bibitem[\protect\citeauthoryear{Buson et al.}{1993}]{Busonetal93} Buson, L., Salder, E., Zeilinger, W., Bertin, G., Bertola, F. et al 1993, A\&A, 280, 409

\bibitem[\protect\citeauthoryear{Buta et al.}{2015}]{Buta15} Buta, R.J.,  Sheth, K., Athanassoula, E.
Bosma, A., Knapen J.H. et al. 2015, ApJS, 217, 32

\bibitem[\protect\citeauthoryear{Capaccioli, Caon \& Rampazzo}{1990}]{Capaccioli90} Capaccioli, M., Caon, N.
Rampazzo, R. 1990, MNRAS, 242, 24p.

\bibitem[\protect\citeauthoryear{Caon, Capaccioli \& D'Onofrio}{1993}]{Caon93} Caon, N.,
Capaccioli, M., D'Onofrio, M. 1993, MNRAS 265, 1013

\bibitem[\protect\citeauthoryear{Capaccioli,  Piotto \& Rampazzo}{1988}]{CPR88} 
Capaccioli,M., Piotto, G., Rampazzo, R. 1988, AJ 96, 487  

\bibitem[\protect\citeauthoryear{Comer\'on et al.}{2014}]{Comeron14} Comer\'on, S. Salo, H.,
Laurikainen, E., Knapen, J.H., Buta, R. et al., 2014, A\&A, 562, A121

\bibitem[\protect\citeauthoryear{Carollo \& Danziger}{1994}]{CD94} Carollo, M., Danziger, I.J. 1994, MNRAS 270, 523

\bibitem[\protect\citeauthoryear{Citterio et al.}{1994}]{swift2}
Citterio, O., Conconi, P., Ghigo, M., et al. 1994, in SPIE Conf. Ser., 
ed. R. B. Hoover, \& A. B. Walker, 2279, 480

\bibitem[\protect\citeauthoryear{Demoulin-Ulrich  et al.}{1984}]{U84}Ulrich-Demoulin M-H., Butcher, H. R., Boksenberg, A., 1984,  ApJ, 285, 527

\bibitem[\protect\citeauthoryear{de Vaucouleurs}{1948}]{deVauc48} de Vaucouleurs, G.
1948, Ann. Astrophys. , 11, 247

\bibitem[\protect\citeauthoryear{de Vaucouleurs et al.}{1991}]{deVauc91} de Vaucouleurs, G., de Vaucouleurs, A., Corwin, H. G., Jr., Buta, R. J., Paturel, G., Fouqu\'e, P. 1991, Third Reference Catalogue of Bright Galaxies. Springer, New York, NY (USA): RC3

\bibitem[\protect\citeauthoryear{Duc et al.}{2015}]{Duc15} Duc, P-A., Cuilllandre, G-C., Karabak, E. Cappellari, M.,
Alatalo, K. et al. 2015, MNRAS, 446, 120.


\bibitem[\protect\citeauthoryear{Erwin et al.}{2015}]{Erwin15} Erwin, P., Saglia, R.P., Fabricius, M., Thomas, J.,
Nowak, N. et al., 2015, MNRAS, 446, 4039

\bibitem[\protect\citeauthoryear{Gadotti}{2009}]{Gadotti09} Gadotti, D. A., 2009, MNRAS, 393, 1531


\bibitem[\protect\citeauthoryear{Gehrels et al.}{2004}]{swift1}
Gehrels, N., Chincarini, G., Giommi, P., et al. 2004, ApJ, 611, 1005

\bibitem[\protect\citeauthoryear{Goudfrooij}{1994}]{Gou94} Goudfrooij, P. 
1994, Ph.D. thesis, University of Amsterdam, The Netherlands

\bibitem[\protect\citeauthoryear{Governato, Reduzzi \& Rampazzo}{1993}]{Governato93} Governato, F., Reduzzi, L.,
Rampazzo, R. 1996, MNRAS, 261, 379

\bibitem[\protect\citeauthoryear{Hern\'andez-P\'erez \& Bruzual}{2014}]{Hernandez14} Hern\'andez-P\'erez, F., Bruzual, G. 2014, MNRAS, 444, 2571

\bibitem[\protect\citeauthoryear{Ho et al.}{2011}]{Ho11} Ho, L.C., Zhao-Yu, L., Barth, A.J. et al. 
2011, ApJS, 197, 21: CGS

\bibitem[\protect\citeauthoryear{Hodges-Kluck \& Bregman}{2014}]{Hodges2014} Hodges-Kluck, E.,
Bregman, J.N. 2014, AJ, 789, 131

\bibitem[\protect\citeauthoryear{Hopkins et al. }{2009}]{Hopkins09} Hopkins, P.F., Cox, T.J.,
Dutta, S.N., Hernquist, L., Kormedy, J., Lauer, T.R. 2099, ApJS, 181, 135

\bibitem[\protect\citeauthoryear{Hoversten et al.}{2011}]{Hoversten11}
Hoversten E.A.,Gronwall, C., Vanden Berk, D.E., Basu-Zych, A. R., Breeveld, A. A. et al. 2011, AJ, 141, 205

\bibitem[\protect\citeauthoryear{Huang et al.}{2013}]{Huang13} Huang, S., Ho, L.C.,  Peng, C.Y., Li, Z-Y., Barth, A.J.  2013, ApJ, 766, 47

\bibitem[\protect\citeauthoryear{Jedrzejewski}{1987}]{Jedr87} Jedrzejewski, R., 1987, MNRAS, 226, 747

\bibitem[\protect\citeauthoryear{Jeong}{2009}]{Jeong09} Jeong, H., Sukyoung K., Bureau, M., Davies, R.L., Falc\'on-Barroso, J.  et al. 2009, MNRAS, 398, 2028

\bibitem[\protect\citeauthoryear{J\'ozsa et al.}{2009}]{Jozsa09} J\'ozsa, G.I.G., Oosterloo, T.A., 
Morganti, R. et al. 2009, A\&A, 494, 489

\bibitem[\protect\citeauthoryear{Kaneda et al.}{2008}]{Kaneda08} Kaneda, H., Honaka, T., Sakon, 
L. et al. 2008, ApJ, 684, 270

\bibitem[\protect\citeauthoryear{Karczewski et al.}{2013}]{Karczewski13}Karczewski, O.\L{}.,  Barlow,  M. J.,   Page, M. J.,  Kuin, N. P. M.,  Ferreras, I. et al. 2013, MNRAS, 431, 2493

\bibitem[\protect\citeauthoryear{Kaviraj et al.}{2007}]{Kaviraj07} Kaviraj S., Schawinski, K., Devriendt, J. E. G., Ferreras, I.,
 Khochfar, S. et al., 2007, ApJS, 173, 619

\bibitem[\protect\citeauthoryear{Kennedy et al.}{2016}]{Kennedy16} Kennedy, R., Bamford, S.P., H\"au\ss ler, B.,
Baldry, I., Bremer, M. et al. 2016, MNRAS, 460, 3458

\bibitem[\protect\citeauthoryear{Kilborn et al.}{2009}]{Kilborn09} Kilborn, V. A., Forbes, D. A., Barnes, D. G., Koribalski, B. S., et al. 2009, MNRAS, 400, 1962


\bibitem[\protect\citeauthoryear{Kim et al.}{1988}]{Kim88} Kim, D.-W., Guhathakurta, P., Van Gorkom, J.H. et al. 1988, ApJ 330, 684
	

\bibitem[\protect\citeauthoryear{La Barbera et al.}{2010}]{LaBarbera2010} La Berbera, F., de Cravalho, R.R., de la Rosa , I.G., Lopes, P.A.A., Kohl-Moreira, L. L., Capelato, H.V. 2010, MNRAS, 408, 1313

\bibitem[\protect\citeauthoryear{Laurikainen et al.}{2006}]{Laurikainen06} Laurikainen, E., Salo, H., Buta, R., Knapen, J., Speltincx, T., Block, D. 2006, AJ, 132, 2634

\bibitem[\protect\citeauthoryear{Laurikainen et al.}{2010}]{Laurikainen10}
Laurikainen, E., Salo, H., Buta, R., Kanpen, J.H., Comer\'on, S.  2010, MNRAS, 405, 1089 

\bibitem[\protect\citeauthoryear{Laurikainen et al.}{2011}]{Laurikainen11}
Laurikainen, E., Salo, H., Buta, R., Kanpen, J.H. 2011, MNRAS, 418, 1452 

\bibitem[\protect\citeauthoryear{Li et al.}{2011}]{Li11} Li, Z.-Y., Ho, L. C., Barth, A. J., Peng, C. 
Y. 2011, ApJS, 197, 22

\bibitem[\protect\citeauthoryear{Longhetti et al.}{2000}]{Longhetti00} Longhetti, M., Bressan, A., Chiosi, C., Rampazzo, R. 2000, A\&A, 353, 917

\bibitem[\protect\citeauthoryear{Makarov et al.}{2014}]{Makarov2014} {{Makarov}, D., {Prugniel}, P., {Terekhova}, N., {Courtois}, H., 	{Vauglin}, I.} 2014, A\&A, 570, A13

\bibitem[\protect\citeauthoryear{ Marino et al.}{2009}]{Marino09} Marino, A., Iodice, E., Tantalo, R., Piovan, L. Bettoni, D. et al. 2009, A\&A, 508, 1235

\bibitem[\protect\citeauthoryear{Marino et al.}{2011a}]{Marino11a} Marino, A., Rampazzo, 
R., Bianchi, L., Annibali, F., Bressan, A., Buson, L. M., Clemens, M. S., Panuzzo, P., Zeilinger, W. W. 2011a, MNRAS, 411, 311 

\bibitem[\protect\citeauthoryear{Marino et al.}{2011b}]{Marino11b} Marino, A., Bianchi, L.,
Rampazzo, R., Thilker, D., Annibali, F., Bressan, A., Buson, L. M. 2011b, Ap\&SS, 335, 243 

\bibitem[\protect\citeauthoryear{Marino et al.}{2011c}]{Marino11c} Marino, A., Bianchi, L., Rampazzo, R.,
Thilker, D. A., Annibali, F., Bressan, A., Buson, L. M. 2011c,  ApJ, 736, 154  

\bibitem[\protect\citeauthoryear{Markwardt et al.}{2009}]{Markwardt2009} Markwardt, C.~B.\ 2009,
Astronomical Data Analysis Software and Systems XVIII, 411, 251

\bibitem[\protect\citeauthoryear{Mapelli et al.}{2015}]{Mapelli15} Mapelli, M., Rampazzo, R., Marino, A. 2015, A\&A,
575, A16

\bibitem[\protect\citeauthoryear{Martin et al.}{2005}]{Martin05} Martin D. C., Fanson, J., Schiminovich, D., Morrissey, P., 
Friedman, P. G. et al., 2005, ApJ, 619, L1

\bibitem[\protect\citeauthoryear{Mazzei et al.}{2014a}]{Mazzei14a} Mazzei, P., Marino, A., Rampazzo, R. 2014a, ApJ, 782, 53

\bibitem[\protect\citeauthoryear{Mazzei et al.}{2014b}]{Mazzei14b} Mazzei, P., Marino, A., Rampazzo, R., Galletta, G., Bettoni, D. 2014b, Advances in Space Research, 53, 950.

\bibitem[\protect\citeauthoryear{Morelli et al.}{2008}]{morelli} Morelli, L.; Pompei, E.; Pizzella, A.; M\'endez$-$Abreu, J.;  Corsini, E. M.; Coccato, L.; Saglia, R. P.; Sarzi, M.; Bertola, F., 2008, MNRAS 389, 341 

 
 \bibitem[\protect\citeauthoryear{Nanni et al.}{2013}]{Nanni13} Nanni, A., Bressan, A. Marigo, P., Girardi, L. 2013, MNRAS, 434, 2390

\bibitem[\protect\citeauthoryear{Oke}{1974}]{Oke74} Oke 1974, ApJS, 27, 21

 \bibitem[\protect\citeauthoryear{Panuzzo et al.}{2011}]{Panuzzo11} Panuzzo, P., Rampazzo, R., Bressan, A., Vega, O., Annibali F., Buson, L.M., Clemens, M.S., Zeilinger, W.W. 2011, A\&A, 528, A10

\bibitem[\protect\citeauthoryear{Pizzella et al.}{1997}]{Pi97} Pizzella, A., Amico, P., Bertola, F. et al. 1997, AA 323, 349


\bibitem[\protect\citeauthoryear{Poole et al.}{2008}]{Poole2008} Poole, T. S., Breeveld, A. A., Page, M. J., Landsman, W., Holland, S. T., et al. 2008, MNRAS, 383, 627
 
 \bibitem[\protect\citeauthoryear{Rampazzo et al.}{2007}]{Rampazzo07} Rampazzo, R., Marino, A., Tantalo, R.,
 Bettoni, D., Buson, L.M. et al. 2007, MNRAS, 381, 245
 
 \bibitem[\protect\citeauthoryear{Rampazzo et al.}{2011}]{Rampazzo11} Rampazzo, R., Annibali, F., Marino, A., Bianchi,
 L., Bressan, A. et al. 2011, Astrphys. Space Scince, 335, 201 

\bibitem[\protect\citeauthoryear{Rampazzo et al.}{2013}]{Rampazzo13} Rampazzo, R., Panuzzo, P., Vega , O. 
et al. 2013, MNRAS, 432, 374

\bibitem[\protect\citeauthoryear{Roming et al.}{2005}]{Roming05} Roming, P.W. A., Kennedy, T. E., Mason, K. O., Nousek, J. A., Ahr, L. et al. 2005, Space Science Rev., 120, 95

\bibitem[\protect\citeauthoryear{Roming et al.}{2009}]{Roming09} Roming, P. W. A., Koch, T. S., Oates, S. R., Porterfield, B. L., Vanden Berk, D. E.,  et al. 2009, ApJ, 690, 163

\bibitem[\protect\citeauthoryear{Ryan-Weber,   Webster \& Starvely-Smith}{2003}]{Ryan-Weber2003} Ryan-Weber, E., Webster, R.,  Starvely-Smith, L.. 2003, MNRAS, 343, 1195


\bibitem[\protect\citeauthoryear{Salim \& Rich}{2010}]{Salim10} Salim, S., Rich, R.M. 2010, ApJ, 714, L290

\bibitem[\protect\citeauthoryear{Salim et al.}{2012}]{Salim12} Salim, S., Fang, J.J., Rich, R.M., Faber, S.M., Thilker, D.A. 2012, ApJ, 755, 105

\bibitem[\protect\citeauthoryear{Salo et al.}{2015}]{Salo2015} Salo, H., Laurikainen, E., Laine, J., Comer\'on, S., Gadotti, D. A.,
et al. 2015, ApJS, 219, 4 

\bibitem[\protect\citeauthoryear{Sandage \& Tammann}{1987}]{RSA} Sandage, A.R., Tammann, G. 1987,  {\it A Revised Shapley Ames Catalogue of Bright Galaxies}, Carnegie, Washington (RSA)

\bibitem[\protect\citeauthoryear{Schawinski et al.}{2007}]{Schawinski07} Schawinski, K., Kaviraj, S., Khochfar, S., Yoon,
S.-J., et al. 2007, ApJS, 173, 512 

\bibitem[\protect\citeauthoryear{Schechter \& Gunn}{1978}]{Schechter78} Schechter, P.L., Gunn, J.E. 1978, AJ, 83, 1360

\bibitem[\protect\citeauthoryear{Schweizer et al.}{1989}]{Schweizer} Schweizer, F.,
van Gorkom, J.H., Seitzer, P. 1989. ApJ 338, 770

\bibitem[\protect\citeauthoryear{Scorza et al.}{1998}]{Scorza98} Scorza, C., Bender, R., Wilkelmann, 
C. et al. 1998, A\&ASS, 131, 265

\bibitem[\protect\citeauthoryear{Scott et al.}{2014}]{Scott14} Scott, N., Davies, R.L., Houghton, 
R.C.W., Cappellari, M., Graham, A.W., Pimbblet, K.A. 2014, MNRAS, 441, 274

\bibitem[\protect\citeauthoryear{Serra \& Oosterloo}{2010}]{Serra10} Serra, P., Oosterloo, T.A. 2010, MNRAS, 401, L29

\bibitem[\protect\citeauthoryear{S\'ersic}{1968}]{Sersic68} S\'ersic, J. L. (ed.) 1968, Atlas de Galaxias Australes (Cordoba, Argentina: Observatorio Astronomico)

\bibitem[\protect\citeauthoryear{Stoughton et al.}{2002}]{Stoughton02} Stoughton C., Lupton, R. H.; Bernardi, M., Blanton, M. R., Burles, S. et al., 2002, AJ, 123, 485

\bibitem[\protect\citeauthoryear{Tal et al.}{2009}]{Tal09} Tal, T. van Dokkum, P.G., Nelan, J., Bezanson,  R. 2009, AJ,  138 1417

\bibitem[\protect\citeauthoryear{Thilker et al.}{2010}]{Thilker10} Thilker, D.A., Bianchi, L., Schiminovic, D. et al. 2010, ApJ, 714, L171

\bibitem[\protect\citeauthoryear{Trinchieri et al.}{2015}]{Trinchieri15} Trinchieri, G., Rampazzo, R.,  Mazzei, P.
Marino, A., Wolter, A. 2015, MNRAS, 449, 3021 

\bibitem[\protect\citeauthoryear{Tully}{1988}]{T88} Tully, R. B. 1988, Nearby Galaxy Catalog, Cambridge University Press

\bibitem[\protect\citeauthoryear{Tully et al.}{2009}]{Tully09} Tully, R. B., Rizzi, L., Shaya, E. J. Courtois, H. M., Makarov, D. I., Jacobs, B. A. 2009, AJ, 138, 323

\bibitem[\protect\citeauthoryear{Ulrich}{1975}]{Ulrich75} Ulrich, M-H. 1975, PASP, 87, 965

\bibitem[\protect\citeauthoryear{Vega et al.}{2010}]{Vega10} Vega, O., Bressan, A., Panuzzo, P., Rampazzo, R., Clemens, M.S. et al. 2010, ApJ, 721, 1090

\bibitem[\protect\citeauthoryear{Vulcani et al.}{2014}]{Vulcani14} Vulcani, B., Bamford, S.P.,
Ha\"u\ss{}ler, B., Vika, M., Rojas A. et al. 2014,  MNRAS, 441, 1340

\bibitem[\protect\citeauthoryear{Yi et al.}{2005}]{Yi05} Yi S. K., Yoon, S.-J., Kaviraj, S., Deharveng, J.-M., Rich, R. M.  et al., 2005, ApJ, 619, L111

\bibitem[\protect\citeauthoryear{Yi et al.}{2011}]{Yi11} Yi S. K., Lee J., Sheen Y.-K., Jeong H., Suh H., Oh K., 2011, ApJS, 195, 22

\bibitem[Werk et al. (2010)]{Werk10} Werk, J. K., Putman, M. E., Meurer, G. R., Ryan-Weber, E. V., Kehrig, C., et al. 2010, AJ, 139, 279

\bibitem[\protect\citeauthoryear{Whitmore et al.}{1990}]{Whitmore90} Whitmore, B.C., Lucas, R.A., 
McElroy, D.B., Steiman-Cameron, T, Y., Sackett, P. D., Olling, R. P. 1990, AJ, 100, 1489

\bibitem[Wyder et al. (2007)]{Wyder07} Wyder, T.K., Martin, D.C., Schiminovich, D., Seibert, M., Bud\'avari, T. et al. 2007,
ApJS, 173, 293  

\bibitem[\protect\citeauthoryear{Zeilinger et al.}{1996}]{Z96} Zeilinger, W.W., Pizzella, A., Amico, P., Bertin, 
G. et al., 1996, AAS 120, 257

\end{thebibliography}
\end{document}